%% file: main.tex
\documentclass[11pt]{article}
\usepackage{verbatim}  
\usepackage{bm, commath}
\usepackage{natbib}
\usepackage{mdframed}
\usepackage{caption}
\usepackage{graphicx}
\usepackage{subfig}
\usepackage{amsmath, amsfonts, amsthm}
\usepackage{float}
\usepackage{booktabs,siunitx}
\usepackage{url}
\usepackage{multirow}
\usepackage{amssymb}
\usepackage{blkarray}
\usepackage{bbm}
\usepackage{multicol}
\usepackage{authblk}
\usepackage[shortlabels]{enumitem}
\usepackage{multirow}

\usepackage{listings}
\usepackage{bbm}
\usepackage{textcomp}
\usepackage[margin=1in]{geometry}
\usepackage{authblk}
\usepackage[ruled]{algorithm2e}
\SetKwInput{KwParam}{Parameter}
\SetAlgoCaptionLayout{centerline}
\usepackage{hyperref}
\usepackage{sectsty}
\usepackage[doublespacing]{setspace}
\usepackage[utf8]{inputenc}
\usepackage{float}
\usepackage{tikz}
\usetikzlibrary{shapes,decorations,arrows,calc,arrows.meta,fit,positioning}

\def\bs{\boldsymbol}

\usepackage[textsize=scriptsize, textwidth = 2cm, shadow]{todonotes}

\newtheorem{theorem}{Theorem}

\newtheorem{remark}{Remark}

\newtheorem*{assumption*}{Assumption}

\newtheorem{proposition}{Proposition}

\newtheorem{innercustomgeneric}{\customgenericname}
\providecommand{\customgenericname}{}
\newcommand{\newcustomtheorem}[2]{%
  \newenvironment{#1}[1]
  {%
   \renewcommand\customgenericname{#2}%
   \renewcommand\theinnercustomgeneric{##1}%
   \innercustomgeneric
  }
  {\endinnercustomgeneric}
}

\newcustomtheorem{customthm}{Theorem}
\newcustomtheorem{customassumption}{Assumption}

\usepackage{mathtools}

\usepackage{xcolor}

\makeatletter
\renewcommand{\algocf@captiontext}[2]{#1\algocf@typo. \AlCapFnt{}#2} 
\def\@algocf@capt@plain{top}
\renewcommand{\algocf@makecaption}[2]{%
  \addtolength{\hsize}{\algomargin}%
  \sbox\@tempboxa{\algocf@captiontext{#1}{#2}}%
  \ifdim\wd\@tempboxa >\hsize
  \hskip .5\algomargin%
  \parbox[t]{\hsize}{\algocf@captiontext{#1}{#2}}
  \else%
  \global\@minipagefalse%
  \hbox to\hsize{\box\@tempboxa}
  \fi%
  \addtolength{\hsize}{-\algomargin}%
}
\makeatother

\allowdisplaybreaks

\begin{document}

\sectionfont{\bfseries\large\sffamily}%

\subsectionfont{\bfseries\sffamily\normalsize}%




\title{Dealing with positivity violations in mediation analysis via weighted controlled effects, with application to assessing immune correlates of protection in antigen-experienced participants}

\author[1]{Qijia He}
\author[2]{Bo Zhang\thanks{Email: {\tt bzhang3@fredhutch.org}} }

\affil[1]{Department of Statistics, University of Washington}
\affil[2]{Vaccine and Infectious Disease Division, Fred Hutchinson Cancer Center}

\date{}

\maketitle

\noindent
\textsf{{\bf Abstract}: \input{abstract}}%

\vspace{0.3 cm}

\input{1.Introduction}
\input{2.Notation}
\input{3.Estimation}

\input{4.Simulation}

\input{5.Case_study}

\input{6.Discussion}

\section*{Acknowledgements}
We thank the participants, site staff, and investigators of the COVAIL trial sponsored by the Division of Microbiology and Infectious Diseases at the National Institute Of Allergy and Infectious Diseases of the National Institutes of Health (NIAID NIH). We thank Dr. Zach Branson, Dr. Peter B. Gilbert, and Dr. Youyi Fong for helpful suggestions. This work is funded by the NIAID NIH (R01AI192632 to BZ). The content is solely the responsibility of the authors and does not necessarily represent the official views of the National Institutes of Health.

\bibliographystyle{apalike}
\bibliography{paper-ref}

\input{Appendix}

\end{document}

%% file: abstract.tex
Causal mediation analysis has become an important and increasingly used framework for evaluating candidate immune response biomarkers in vaccine research. A controlled effects approach has been proposed to estimate controlled risk curves under a counterfactual scenario in which the entire study population is vaccinated and their post-vaccination immune responses are set to a range of fixed levels. This framework performs well when the study population is antigenically na\"ive, that is, individuals have not been previously exposed to the antigen, as is common in HIV-1 vaccine research and during the early phases of the COVID-19 pandemic. However, the controlled effects framework becomes more challenging to apply in antigen-experienced populations, where prior vaccination or infection has occurred, as in the case of influenza, dengue, and more recent phases of the COVID-19 pandemic. In such settings, a key identification assumption for valid causal mediation analysis, the positivity assumption, is violated: it is no longer plausible to conceive of a hypothetical intervention that sets a post-vaccination immune marker to a fixed level below an individual’s baseline immune level. In this article, we introduce a weighted controlled risk approach that targets a subpopulation for whom there is a prespecified probability of attaining a post-vaccination immune marker level. We further generalize this framework to study contrasts of controlled risks for relevant subpopulations. We demonstrate the validity of the proposed estimators through simulation studies and apply the method to reanalyze post-vaccination neutralizing antibody titers against Omicron BA.4/BA.5 as an immune correlate of COVID-19 in the Coronavirus Variant Immunologic Landscape (COVAIL) trial. \textsf{R} code to implement the proposed method can be found on Github: \url{https://github.com/Qijia-He/weighted_CVE}.

%% file: 1.Introduction.tex
\section{Introduction}
\label{sec: intro}

\subsection{Vaccine development; surrogate endpoint; statistical framework}
\label{subsec: Vaccine development}
Surrogate endpoints have received considerable attention across many areas of medicine. An ideal surrogate endpoint provides early evidence of treatment efficacy, thereby reducing the need to wait for definitive clinical outcomes that are often observed only after lengthy and logistically challenging clinical trials \citep{joffe2009related}. In vaccine development, an immune correlate of risk (CoR) is an immunologic biomarker that predicts a clinical endpoint among vaccine recipients \citep{qin2007framework}. An immune correlate of protection (CoP) is a CoR that reliably predicts vaccine efficacy (VE) relative to a placebo, or relative vaccine efficacy (relVE) compared with another vaccine \citep{qin2007framework}. Candidate CoRs and CoPs include immune responses such as neutralizing antibody levels, binding antibody levels, and T cell responses, typically measured at a prespecified, approximately peak time point (e.g., $2$ to $4$ weeks post-vaccination). Identifying and validating immune correlates generally requires data from multiple phase~3, placebo-controlled trials. Once validated, a CoP may be used to support approval of a modified vaccine formulation or to extend licensure to new populations not originally included in the efficacy trials \citep{gilbert2022covid}.

Four major statistical frameworks have been developed for evaluating surrogate endpoints: the Prentice criteria \citep{prentice1989surrogate}, principal stratification \citep{frangakis2002principal,gilbert2008evaluating}, meta-analytic approaches \citep{daniels1997meta,molenberghs2008meta}, and causal mediation analysis \citep{albert2008mediation,cowling2019influenza,hejazi2021efficient,gilbert2023controlled}. \citet{prentice1989surrogate} introduced one of the earliest formal frameworks and proposed a set of conditions under which rejection of the null hypothesis of no treatment effect on a surrogate endpoint implies rejection of the null hypothesis of no treatment effect on the true clinical outcome.  

The principal stratification framework, proposed by \citet{frangakis2002principal}, evaluates treatment effects within principal strata defined by the joint potential surrogate outcomes under both treatment assignments; see \citet{gilbert2008evaluating} for extensions to continuous surrogate outcomes and applications to vaccine trials. In contrast, the meta-analytic approach assesses surrogacy by examining the association between treatment effects on the surrogate and on the clinical outcome across multiple studies, enabling validation of surrogacy at the study level \citep{daniels1997meta,molenberghs2008meta,stijven2025evaluation}.

More recently, causal mediation analysis has assumed an increasingly prominent role in immune correlates research. \citet{gilbert2023controlled} introduced a controlled effects framework for evaluating surrogate endpoints, centered on two key estimands: controlled risk and controlled vaccine efficacy. Controlled risk is defined as the risk of a clinical endpoint under a counterfactual scenario in which all individuals are vaccinated and their immune marker levels are set to a prespecified value. Controlled vaccine efficacy compares this controlled risk under vaccination with the corresponding placebo risk, while controlled relative vaccine efficacy contrasts controlled risks for two vaccines, each evaluated at a specified immune marker level. This framework has been widely applied in recent immune correlates analyses of COVID-19 vaccines \citep{gilbert2022covid,gilbert2022immune,fong2022immune,fong2023immune,zhang2024omicron,zhang2025neutralizing,mkhize2025neutralizing} and serves as the conceptual foundation for the present article.

An alternative to the controlled effects framework is the modified treatment policy (MTP) or stochastic intervention approach \citep{munoz2012population,haneuse2013estimation,hejazi2021efficient}, which evaluates the risk of a clinical endpoint under hypothetical interventions that modify the distribution of an immune marker relative to its observed distribution. A commonly studied example is a rightward shift of the observed immune marker distribution, corresponding to a uniform increase in immune marker levels by a specified amount \citep{huang2023stochastic}. Under this formulation, the stochastic intervention framework addresses the question: how would the risk of the clinical endpoint change if the vaccine elicited immune responses that were uniformly higher than the observed levels?

\subsection{Violation of the positivity assumption in the controlled effects approach}
\label{subsec: Non-naive population}
In many vaccine efficacy trials, including those for HIV-1 and SARS-CoV-2 during the early phase of the COVID-19 pandemic, the study population consisted of individuals with no prior exposure to the pathogen and no vaccination (i.e., antigen-na\"ive or na\"ive). Among these na\"ive participants, there is structural knowledge that no baseline antigen-specific immune response is present. Accordingly, the baseline immune marker can be assigned a ``negative” or lowest possible value, or classified as ``no response,” corresponding to measurements below the assay’s limit of detection (LOD). In this article, we denote this as “$<\textsf{LOD}$.”

As the pandemic progressed, however, vaccine trials, particularly those evaluating the booster vaccines, started to enroll non-na\"ive participants, who had diverse immune histories due to prior infection, vaccination, or both. For non-na\"ive participants, baseline immune marker levels can influence both the post-vaccination immune response and subsequent clinical outcomes.

A key assumption in controlled effects mediation analysis is the positivity assumption, which requires that, conditional on baseline covariates, each individual has a non-zero probability of receiving any level of the mediator (see, e.g. \citealp{joffe2009related}). Among non-na\"ive participants, this assumption is often violated, due to the typically unidirectional relationship between baseline and post-vaccination immune responses---that is, the post-vaccination immune marker level is unlikely to fall below its baseline level. 

One strategy to address this challenge is to adopt the stochastic intervention approach and focus on estimating the risk of clinical endpoint under hypothetical shifts of the mediator to higher (or lower) values relative to those observed. For example, \citet{hejazi2021efficient} employed this approach to estimate the counterfactual probability of HIV-1 infection under stochastic interventions that increased the standardized CD4+ or CD8+ polyfunctionality score by varying degrees (e.g., half of a standard deviation). \citet{huang2023stochastic} applied this approach to study neutralizing and binding antibody markers as correlates against symptomatic COVID-19 in the COVE mRNA-1273 Trial. 

Outside the stochastic intervention framework, a common strategy for addressing positivity violation is to trim extreme propensity score estimates \citep{heckman1997matching, frolich2004programme, crump2009dealing, yang2018asymptotic, branson2023causal, song2025generative}. In recent work, \citet{branson2023causal} rigorously studied propensity score trimming in settings with a continuous point exposure. \citet{branson2023causal} proposed efficient influence function-based estimators for trimmed average treatment effect and studied their theoretical properties. 

In this paper, we generalize the approach of \citeauthor{branson2023causal} \citeyearpar{branson2023causal} to a causal mediation analysis setting in which the positivity assumption is violated for the mediator, due to reasons described above. Specifically, within the controlled effects framework, we modify the controlled effects estimands for evaluating correlates of risk and correlates of relative vaccine efficacy in non-na\"ive study populations.

The remainder of the article is organized as follows. In Section~\ref{sec: notation and estimand}, we introduce notation and define the causal estimands of interest. Section~\ref{sec: estimation} derives the efficient influence function and presents the proposed estimators. The finite-sample performance of these estimators is investigated through simulation studies in Section~\ref{sec: Simu}. In Section~\ref{sec: case study}, we apply the proposed methods to reanalyze neutralizing antibody levels as immune correlates in the Coronavirus Variant Immunologic Landscape (COVAIL) trial \citep{branche2023comparison,branche2023immunogenicity,fong2025neutralizing,zhang2025neutralizing}, a study of the second COVID-19 booster vaccine. Section~\ref{sec: discussion} concludes with a discussion.

%% file: 2.Notation.tex
\section{Notation and estimand}
\label{sec: notation and estimand}
\subsection{Controlled risk among na\"ive participants}
\label{subsec: CoR naive}
Let $B$ denote the baseline level of an immune marker, 
$A$ the vaccine assignment, and $S$ the same immune marker measured at a prespecified post-vaccination study visit. For example, in the COVAIL trial, $A$ represents receipt of a second COVID-19 booster dose; $B$ corresponds to the baseline neutralizing antibody ID\textsubscript{50} titer (nAb-ID\textsubscript{50}) against a viral strain of interest (e.g., Omicron BA.4/5); and $S$ denotes the approximate peak nAb-ID\textsubscript{50} level measured 15 days post-vaccination. 

In addition, let $\bs{X}$ denote a vector of baseline covariates, including known risk factors, and let $Y$ denote a clinical endpoint of interest. In this paper, we will focus on a continuous or binary clinical endpoint. Following the potential outcomes (PO) framework \citep{neyman1923application,rubin1974estimating}, we use $Y(A = a, S = s)$ to denote the potential outcome that would have been observed had an individual received treatment $A=a$ and had their post-vaccination immune marker level set to $S = s$. The potential outcome $Y(A = a, S = s)$ is different from $Y(A = a, S(A = a))$ as the former defines a potential outcome under a joint intervention of $A = a$ and $S = s$, while the latter only intervenes on $A$ and sets the post-vaccination marker at its natural level $S(A = a)$ under treatment $A = a$.

We follow \citet{gilbert2023controlled} and define the following conditional controlled risk:
\begin{equation*}
    r(a, s, B, \bs{X}) := r(A = a, S = s \mid B, \bs{X}) = \mathbb{E}[Y(A = a, S = s) \mid B, \bs{X}],
\end{equation*}
which is the mean counterfactual outcome within strata defined by the baseline immune marker $B$ and covariates $\bs{X}$. Among na\"ive study participants, the baseline immune marker $B$ typically falls below the limit of detection (LOD) and set as such. In a series of correlates studies for COVID-19 among baseline na\"ive study participants during the early phase of the pandemic \citep{gilbert2022immune, fong2022immune,fong2023immune,benkeser2023immune}, the causal parameter of interest is $\mathbb{E}_{\mathcal{P}_{\bs{X}}}[r(a, s, B < \textsf{LOD}, \bs{X})]$, where the expectation is taken with respect to $\mathcal{P}_{\bs{X}}$, the marginal distribution of $\bs{X}$, and the baseline marker level $B$ is irrelevant in these studies as $B < \textsf{LOD}$ with probability $1$. 

Two assumptions are key to identifying the controlled risk parameter,  $\mathbb{E}_{\mathcal{P}_{\bs{X}}}[r(a, s, B < \textsf{LOD}, \bs{X})]$, from the observed data: sequential ignorability and positivity \citep{joffe2009related,gilbert2023controlled}. The sequential ignorability assumption essentially says that $S$ is randomized within strata defined by $\bs{X}$. This assumption is often relaxed in sensitivity analyses \citep{gilbert2023controlled}. The positivity assumption requires that $P(S = s \mid A = a, B < \textsf{LOD}, \bs{X} = \bs{x}) > 0$ almost surely, which is mostly reasonable for na\"ive participants. Under sequential ignorability, positivity, and randomization of $A$ (which is often satisfied by study design), the controlled risk parameter can be identified from the observed data as
$\mathbb{E}_{\mathcal{P}_{\bs{X}}}\left[\mathbb{E}\left[Y \mid A = a, S = s, \bs{X}\right]\right]$ using g-computation; see, e.g., \citet{Robins1986,tchetgen2012semiparametric, gilbert2023controlled}.

\subsection{Weighted controlled risk among non-na\"ive participants}
\label{subsec: CoR non-naive}
Participants enrolled in recent booster vaccine trials and included in immune correlates analyses are typically non-na\"ive, with heterogeneous levels of baseline immunity arising from prior infection, vaccination, or both. Consequently, the baseline immune marker $B$ no longer satisfies $B < \textsf{LOD}$ with probability 1. Two questions arise. First, can $B$ be simply ignored? Based on current understanding of immune dynamics, the baseline immune marker $B$ is likely to influence both the peak post-vaccination immune marker $S$ and the clinical outcome $Y$. As a result, the conditional independence assumption $Y(a, s) \perp S \mid B, \bs{X}$ is more plausible than $Y(a, s) \perp S \mid \bs{X}$. Without accounting for $B$, the parameter $\mathbb{E}_{\mathcal{P}_{\bs{X}}}[\mathbb{E}[Y \mid A = a, S = s, \bs{X}]]$ remains a well-defined statistical parameter and can be interpreted as covariate-$\bs{X}$-adjusted, peak immune marker-specific risk in the arm $A = a$. 

Second, which baseline immune responses should be included in $B$? In the COVID-19 context, participants who were previously vaccinated but not infected would typically exhibit baseline immune responses to the Spike protein encoded by the vaccine. In contrast, participants with prior infection would also show immune responses to additional viral proteins, such as nucleocapsid (N) and membrane (M). Ideally, $B$ should incorporate responses to Spike as well as N and/or M to more comprehensively characterize baseline immunity and to stratify participants into subgroups defined by distinct immunologic histories. Furthermore, immune responses elicited by prior vaccination versus natural infection may differ not only in magnitude but also in quality; for instance, infection may induce B and T cells with greater functional capacity. For confounding control, this suggests that, when available, researchers should include variables capturing immune response quality, such as antibody affinity or avidity, alongside measures of response magnitude.

We study how to define controlled effects-style estimands for baseline non-na\"ive participants. We begin by studying how to evaluate the mean potential outcome $Y(a, s)$ within a ``relevant" subpopulation of the study population---specifically, non-na\"ive participants who have at least probability $t$ of achieving $S = s$ if given vaccine $A = a$. This subpopulation is considered ``relevant" for vaccine arm $A = a$ and peak immune marker level $S = s$ because it is conceivable that a vaccine could elicit an immune marker level as high as $s$ in these individuals. To illustrate this, consider a simple example: let $B \sim N(0, 1)$ and define $S = B + \text{Gamma}(2, 1)$. In this setup, only individuals with $B \leq s$ have a positive probability of achieving $S = s$; those with $B > s$ do not, and thus would not be included in the relevant subpopulation for any choice of $t$.

Selecting a relevant subpopulation can be accomplished via weighting. We consider the following weighted controlled risk parameter:
\begin{equation}\label{eq: weighted CR}
   \text{WCR}(a, s) := \frac{\mathbb{E}_{\mathcal{P}_{B, \bs{X}}}\left[\omega_{s}(B, \bs{X}) \cdot r(a, s, B, \bs{X})\right]}{\mathbb{E}_{\mathcal{P}_{B, \bs{X}}}\left[\omega_{s}(B, \bs{X})\right]},
\end{equation}
where $\omega_{s}(B, \bs{X})$ is a weighting scheme that is specific to the vaccine assignment $A = a$ and the post-vaccination immune marker level $S = s$, and is a function of baseline covariates $(B, \bs{X})$. For notational simplicity, we suppress the dependence on $A = a$ in writing $\omega_{s}(B, \bs{X})$. The expectation in \eqref{eq: weighted CR} is taken with respect to $\mathcal{P}_{B, \bs{X}}$, the joint distribution of $(B, \bs{X})$ in the non-na\"ive study population. 

To focus on the controlled risk within the aforementioned ``relevant" subpopulation, we define the weighting function $\omega_{s}(B, \bs{X}) = \omega^{\text{trim}}_{s}(B, \bs{X})$, where
\begin{equation}
    \label{eqn: CoR weight}
    \begin{split}
    \omega^{\text{trim}}_{s}(B, \bs{X}) &= \mathbbm{1}\left\{P(S = s \mid A = a, B, \bs{X}) > t\right\}.
    \end{split}
\end{equation}
\noindent Using this, we define the trimmed weighted controlled risk (TWCR) as
\begin{equation}
    \label{eq: TWCR CoR}
    \text{TWCR}(a, s) = 
\frac{\mathbb{E}_{\mathcal{P}_{B, \bs{X}}}\left[\omega^{\text{trim}}_{s}(B, \bs{X}) \cdot r(a, s, B, \bs{X})\right]}{\mathbb{E}_{\mathcal{P}_{B, \bs{X}}}\left[\omega^{\text{trim}}_{s}(B, \bs{X})\right]} =     \frac{\mathbb{E}_{\mathcal{P}_{B, \bs{X}}}\left[\mathbbm{1}\{\pi(s \mid a, B, \bs{X}) > t \}\cdot r(a, s, B, \bs{X})\right]}{\mathbb{E}_{\mathcal{P}_{B, \bs{X}}}\left[\mathbbm{1}\{\pi(s \mid a, B, \bs{X}) > t \}\right]},
\end{equation}
where $\pi(s \mid a, B, \bs{X})$ denote $P(S = s \mid A = a, B, \bs{X})$, $\mathbbm{1}\{\cdot\}$ denotes the indicator function, and $t > 0$ is a prespecified threshold close to $0$. Estimand like $\text{TWCR}(a, s)$ appears in the literature on propensity score trimming in a point exposure setting \citep{crump2009dealing,yang2018asymptotic,branson2023causal}, where researchers discard subjects whose probabilities of receiving the treatment are below a threshold when estimating causal effects.

As pointed out in \citet{branson2023causal}, the parameter like $\text{TWCR}(a, s)$ is not pathwise differentiable because of (1) the indicator function; and (2) the lack of differentiability at the continuous mediator $S = s$. To overcome the lack of differentiability due to the indicator function, we replace it with a smooth function $\phi\{\pi(s \mid a, B, \bs{X}), t\}$ that approximates the non-differentiable indicator function. One popular choice is to let
$\phi\{\pi(s \mid a, B, \bs{X}), t\} = \Phi_\epsilon\{\pi(s \mid a, B, \bs{X}) - t\}$,
where $\Phi_\epsilon(z)$ denotes the cumulative distribution function (CDF) of a Normal distribution with mean zero and variance $\epsilon^2$; see, e.g., \citet{yang2018asymptotic} and \citet{branson2023causal}. Here, $\epsilon > 0$ controls the extent of approximating the indicator function with the normal CDF. To further circumvent the non-differentiability at the continuous mediator dose level $S = s$, a popular strategy is to consider an integral functional that smooths across the mediator using a kernel function $K_h(\cdot)$ with bandwidth $h$; see, e.g., \citet{branson2023causal}, among others.

After applying these two modifications to the TWCR estimand, we arrive at the following smoothed trimmed weighted controlled risk (STWCR):
\begin{equation}
\label{eq: STWCR}
\text{STWCR}(a, s) =
\frac{\tau_{\omega_{s}^{\text{s-trim}}}^{\text{num}}(a, s)}
{\tau_{\omega_{s}^{\text{s-trim}}}^{\text{den}}(a, s)},
\end{equation}
where
$$
\tau_{\omega_{s}^{\text{s-trim}}}^{\text{num}}(a, s) = \mathbb{E}_{\mathcal{P}_{B, \bs{X}}}\left[\int_{s' \in \mathcal{S}} K_h(s^\prime - s) \phi\left\{\pi(s^\prime \mid a, B, \bs{X}), t\right\}  r(a, s^\prime, B, \bs{X})  ds^\prime \right],$$
$$
\tau_{\omega_{s}^{\text{s-trim}}}^{\text{den}}(a, s) = \mathbb{E}_{\mathcal{P}_{B, \bs{X}}}\left[\int_{s' \in \mathcal{S}} K_h(s^\prime - s) \phi\left\{\pi(s^\prime \mid a, B, \bs{X}), t\right\}  ds^\prime\right],$$
and $\mathcal{S}\subseteq \mathbb{R}$ is the support of $S$.

The parameter $\text{STWCR}(a, s)$ is now pathwise differentiable and asymptotically linear under standard regularity conditions. 

\subsection{Controlled risk correlates of protection among non-na\"ive participants}
\label{subsec: CoP}
A comparison of two controlled risk parameters is referred to as a controlled risk correlate of protection (CoP) analysis \citep{gilbert2023controlled}. For example, in an analysis of na\"ive participants, the role of the peak immune marker $S$ can be evaluated by comparing the controlled risk parameters $\mathbb{E}_{\mathcal{P}_{\bs{X}}}[r(A = a, S = s_0, \bs{X})]$ and $\mathbb{E}_{\mathcal{P}_{\bs{X}}}[r(A = a, S = s_1, \bs{X})]$ within the vaccine arm $A = a$, for different values of $s_0$ and $s_1$. Because both parameters are defined on the same study population (i.e., the entire na\"ive study population), their comparison yields a valid causal contrast. A common visualization is to plot the controlled risk parameter as a function of the peak immune marker $S$; see, for example, \citet{gilbert2022immune,zhang2024omicron,zhang2025neutralizing}.

\citet{gilbert2023controlled} defines controlled relative vaccine efficacy (CRVE) as 
\begin{equation}
    \label{eq: CVE naive}
    \text{CRVE}(s_1, s_0) = 1 - \frac{\mathbb{E}_{\mathcal{P}_{\bs{X}}}[r(A = a_1, S = s_1, \bs{X})]}{\mathbb{E}_{\mathcal{P}_{\bs{X}}}[r(A = a_0, S = s_0, \bs{X})]},
\end{equation}
and it can be interpreted as a contrast in the controlled risk comparing (1) assigning the na\"ive study population to an investigational vaccine $A = a_1$ and setting their peak immune marker $S = s_1$ and (2) assigning the same population to a comparator vaccine $A = a_0$ and setting their peak immune marker $S = s_0$. Again, the CRVE parameter in \eqref{eq: CVE naive} is a well-defined causal contrast as the numerator and denominator are potential outcomes defined on the same na\"ive study population.

To accommodate a non-na\"ive population within the controlled risk CoP framework and to define an appropriate estimand, we again restrict attention to a relevant target population. For example, if the goal is to compare different peak immune response levels, e.g., $S = s_1$ versus $S = s_0$, within the same vaccine arm $A = a$, then the relevant population consists of individuals who have at least probability $t$ of achieving $S = s_1$ after receiving $A = a$, and at least probability $t$ of achieving $S = s_0$ after receiving $A = a$. Analogously, to compare different levels of $S$ across different vaccine arms, the relevant population consists of individuals who have at least probability $t$ of achieving $S = s_1$ after receiving $A = a_1$, and at least probability $t$ of achieving $S = s_0$ after receiving $A = a_0$. In what follows, we focus on the latter setting, noting that the former is a special case obtained by setting $a_0 = a_1$.

Building on the weighted controlled risk estimand developed in Section \ref{subsec: CoR non-naive}, we consider the following trimmed weighted controlled relative vaccine efficacy (TWCRVE):
\begin{equation}\label{eqn: CVE}
    \text{TWCRVE}(a_1, a_0, s_1, s_0) :=  1 - \frac{\mathbb{E}_{\mathcal{P}_{B, \bs{X}}}\left[\omega_{s_1, s_0}^{\text{d-trim}}(B, \bs{X}) \cdot r(a_1, s_1, B, \bs{X})\right]}{\mathbb{E}_{\mathcal{P}_{B, \bs{X}}}\left[\omega_{s_1, s_0}^{\text{d-trim}}(B, \bs{X}) \cdot r(a_0, s_0, B, \bs{X})\right]}.
\end{equation}
In expression \eqref{eqn: CVE}, the weighting function $\omega_{s_1, s_0}^{\text{d-trim}}$ is a double-trimming weight defined as follows:
\begin{equation*}
    \begin{split}
        \omega_{s_1, s_0}^{\text{d-trim}}(B, \bs{X}) := &\mathbbm{1}\{P(S = s_1 \mid A = a_1, B, \bs{X}) > t\}  \times \mathbbm{1}\{P(S = s_0 \mid A = a_0, B, \bs{X}) > t\}, 
    \end{split}
\end{equation*}
for some small $t > 0$.

\begin{remark}\rm
For the non-na\"ive study population, a direct comparison of $\text{TWCR}(a_1, s_1)$ and $\text{TWCR}(a_0, s_0)$ for $s_0 \neq s_1$, as defined in \eqref{eq: TWCR CoR}, does \emph{not} constitute a valid causal comparison because $\text{TWCR}(a_1, s_1)$ and $\text{TWCR}(a_0, s_0)$ represent the mean potential outcomes for different subpopulations, even in the special case where $a_0 = a_1$.
\end{remark}

Similar to the approach used for TWCR, we smooth $\text{TWCRVE}(a_1, a_0, s_1, s_0)$ by replacing both indicator functions in $ \omega_{s_1, s_0}^{\text{d-trim}}(B, \bs{X})$ with smoothed functions and applying kernel smoothing to the continuous mediator levels $S = s_1$ and $S = s_0$. 

A smoothed version of the numerator $\mathbb{E}_{\mathcal{P}_{B, \bs{X}}}\left[\omega_{s_1, s_0}^{\text{d-trim}}(B, \bs{X}) \cdot r(a_1, s_1, B, \bs{X})\right]$ in \eqref{eqn: CVE} is 
\begin{align*}
    \tau^{\text{num}}_{\omega_{s_1, s_0}^{\text{sd-trim}}}(a_1, s_1) := ~&\mathbb{E}_{\mathcal{P}_{B, \bs{X}}}\left[\int_{s' \in \mathcal{S}} K_h(s^\prime - s_0)\phi\left\{{\pi}(s^\prime \mid a_0, B, \bs{X}), t\right\} ds^\prime \right.\\
    &\quad\left. \times \int_{s'' \in \mathcal{S}} K_h(s^{\prime \prime}- s_1)\phi\left\{{\pi}(s^{\prime\prime} \mid a_1, B, \bs{X}), t\right\}r(a_1, s^{\prime\prime}, B, \bs{X}) ds^{\prime\prime}\right],
\end{align*}
and a smoothed version of the denominator $\mathbb{E}_{\mathcal{P}_{B, \bs{X}}}\left[\omega_{s_1, s_0}^{\text{d-trim}}(B, \bs{X}) \cdot r(a_0, s_0, B, \bs{X})\right]$ is
\begin{align*}
   \tau^{\text{den}}_{\omega_{s_1, s_0}^{\text{sd-trim}}}(a_0, s_0) := ~&\mathbb{E}_{\mathcal{P}_{B, \bs{X}}}\left[\int_{s' \in \mathcal{S}} K_h(s^\prime - s_0)\phi\left\{{\pi}(s^\prime \mid a_0, B, \bs{X}), t\right\} r(a_0, s^{\prime}, B, \bs{X})ds^\prime \right.\\
    &\quad\left. \times \int_{s'' \in \mathcal{S}} K_h(s^{\prime \prime}- s_1)\phi\left\{{\pi}(s^{\prime\prime} \mid a_1, B, \bs{X}), t\right\} ds^{\prime\prime}\right].
\end{align*}
\noindent Finally, the smoothed trimmed weighted controlled relative vaccine efficacy, or STWCRVE, can be defined accordingly as follows:
\begin{equation}
\label{eq: STCRVE}
\text{STWCRVE}(a_1, a_0, s_1, s_0) =
1 - \frac{\tau_{\omega_{s_1, s_0}^{\text{sd-trim}}}^{\text{num}}(a_1, s_1)}
{\tau_{\omega_{s_1, s_0}^{\text{sd-trim}}}^{\text{den}}(a_0, s_0)}.
\end{equation}

\subsection{Interpretation of weighted controlled risk and weighted controlled relative vaccine efficacy}
\label{subsec: interpretation}
The parameter $\text{TWCR}(a, s)$ aims at estimating the potential outcome of a subpopulation. The parameter $\text{TWCR}(a, s)$ and its smoothed version $\text{STWCR}(a, s)$ should be interpreted with caution, as the population over which the potential outcome is defined and evaluated varies with both the mediator level $S = s$ and the threshold $t$. While the dependence on the threshold $t$ is primarily technical---differences between, for example, $t = 0.01$ and $t = 0.001$ are typically small and scientifically less consequential---the dependence on the mediator level $s$ may be substantial. The study population relevant to a small value of $s$ may differ meaningfully from that relevant to a large value. For example, when $s$ is small, the ``relevant population” --- defined as the subpopulation of non-na\"ive individuals with a non-trivial probability of attaining a low post-vaccination immune marker level---is expected to have limited baseline immunity. This may occur because participants’ last infection or vaccination was sufficiently remote that acquired immunity has waned, or because they are immunocompromised and unable to mount adequate immune responses following infection or vaccination. In contrast, the relevant population for a large $s$ may include a substantially broader and more heterogeneous subpopulation. 


The trimmed weighted controlled risk is useful in two contexts. First, it serves as a technical construct for deriving the trimmed weighted controlled relative vaccine efficacy. Second, by comparing $\text{TWCR}(a, s)$ to the overall risk in the same subpopulation, that is,
$\frac{\mathbb{E}_{\mathcal{P}_{B, \bs{X}}}\left[\omega^{\text{trim}}_{s}(B, \bs{X}) \cdot r'(a, B, \bs{X}) \right]}{\mathbb{E}_{\mathcal{P}_{B, \bs{X}}}\left[\omega^{\text{trim}}_{s}(B, \bs{X})\right]}$, where $r'(a, B, \bs{X}) = \mathbb{E}[Y \mid A = a, B, \bs{X}],$ researchers can assess whether setting the immune marker level to $s$ reduces the risk of the clinical endpoint for this subpopulation.

Trimmed weighted controlled relative vaccine efficacy is a meaningful causal contrast. For a binary or cumulative incidence endpoint, $\text{TWCRVE}(a_1, a_0, s_1, s_0)$ and its smoothed version quantify the causal relative risk comparing two joint interventions: (1) assigning vaccine $A = a_1$ and peak immune marker $S = s_1$ and (2) assigning vaccine $A = a_0$ and peak immune marker $S = s_0$. This contrast is well-defined over the subpopulation for whom either joint intervention---$(A = a_1, S = s_1)$ or $(A = a_0, S = s_0)$---can be plausibly conceived of. By setting $a_1 = a_0 = a$ in $\text{TWCRVE}(a_1, a_0, s_1, s_0)$, one can estimate $\text{STWCRVE}(a, a, s_1, s_0)$ and investigate whether elevating $S$ from $s_0$ to $s_1$ in the vaccine arm $A = a$ helps decrease the disease risk for the relevant subpopulation, and such an analysis contributes to the controlled risk CoP analysis.

Certain choices of $(s_1, s_0)$ are particularly informative. For instance, setting $s_1 = s_0 = s$ in $\text{STWCRVE}(a_1, a_0, s, s)$ yields the controlled direct effect of vaccine $A = a_1$ compared to $A = a_0$ over a relevant subpopulation. The quantity captures the benefit of vaccine $A = a_1$ relative to $A = a_0$ that is not mediated through $S$. 

%% file: 3.Estimation.tex
\section{Estimation and inference}
\label{sec: estimation}

Proposition \ref{prop: EIF for CoR} derives the efficient influence functions (EIFs) for the numerator $\tau_{\omega_{s}^{\text{s-trim}}}^{\text{num}}(a, s)$ and denominator $\tau_{\omega_{s}^{\text{s-trim}}}^{\text{den}}(a, s)$ of $\text{STWCR}(a, s)$, for fixed $t$ and prespecified $h$ and $\epsilon$.

\begin{proposition}
\label{prop: EIF for CoR}
Let $\mathcal{S}\subseteq \mathbb{R}$ denote the support of $S$. The uncentered efficient influence function for $\tau_{\omega_{s}^{\text{s-trim}}}^{\text{num}}(a, s)$ and $\tau_{\omega_{s}^{\text{s-trim}}}^{\text{den}}(a, s)$ are
{\small
\begin{equation*} 
\centering
\begin{split}
   \varphi^{\text{num}}(a, s) 
   & = \mathbb{IF}\left(\tau_{\omega_{s}^{\text{s-trim}}}^{\text{num}}(a, s)\right) + \tau_{\omega_{s}^{\text{s-trim}}}^{\text{num}}(a, s)\\
&=K_h(S-s) \frac{I(A = a)}{\pi^\prime(a\mid B, \bs{X})}\frac{\partial \phi\left\{\pi\left(S \mid a, B, \bs{X}\right), t\right\}}{\partial \pi\left(S \mid a, B, \bs{X}\right)}\\
& - \int_{s_0\in \mathcal{S}} K_h\left(s_0-s\right) \frac{I(A = a)}{\pi^\prime(a\mid B, \bs{X})} \frac{\partial \phi\left\{\pi\left(s_0 \mid a, B, \bs{X}\right), t\right\}}{\partial \pi\left(s_0 \mid a, B, \bs{X}\right)} \pi\left(s_0 \mid a, B, \bs{X}\right) d s_0 \\
& +\int_{s_0\in \mathcal{S}} K_h\left(s-s_0\right) \phi\left\{\pi\left(s_0 \mid a, B, \bs{X}\right), t\right\}{d s_0},
\end{split}
\end{equation*}
and 
\begin{equation*} 
\centering
\begin{split}
  \varphi^{\text{den}}(a, s) &=
  \mathbb{IF}\left(\tau_{\omega_{s}^{\text{s-trim}}}^{\text{den}}(a, s)\right) + \tau_{\omega_{s}^{\text{s-trim}}}^{\text{den}}(a, s)\\
  &= K_h(S - s) \frac{I(A=a)}{\pi^{\prime}\left(a \mid B, \bs{X}\right)} \frac{\partial \phi\left\{\pi\left(S \mid a, B, \bs{X}\right), t\right\}}{\partial \pi\left(S \mid a, B, \bs{X}\right)} r\left(a, S, B, \bs{X}\right) \\
&+K_h\left(S-s\right)\frac{I(A=a)}{\pi^{\prime}\left(a \mid B, \bs{X}\right)}\frac{ \phi\left\{\pi\left(S \mid a, B, \bs{X}\right), t\right\}}{ \pi\left(S \mid a, B, \bs{X}\right)} \left(Y - r(a, S, B, \bs{X})\right)\\
& -\int_{s_0\in \mathcal{S}} K_h\left(s_0-s\right) \frac{I(A = a)}{\pi^{\prime}\left(a \mid B, \bs{X}\right)} \frac{\partial \phi\left\{\pi\left(s_0 \mid a, B, \bs{X}\right), t\right\}}{\partial \pi\left(s_0 \mid a, B, \bs{X}\right)} \pi\left(s_0 \mid a, B, \bs{X}\right) r\left(a, s_0, B, \bs{X}\right) d s_0\\
&+\int_{s_0\in \mathcal{S}} K_h\left(s-s_0\right) \phi\left\{\pi\left(s_0 \mid a, B, \bs{X}\right), t\right\} r\left(a, s_0, B, \bs{X}\right) d s_0, \\
\end{split}
\end{equation*}
}
respectively, where $\pi'(a \mid B, \bs{X}) = P(A = a \mid B, \bs{X})$, $\pi(s \mid a, B, \bs{X}) = P(S = s \mid A = a, B, \bs{X})$, and $\phi\{\cdot, t\} = \Phi_\epsilon(\cdot; t)$ and $K_h(\cdot)$ are defined in Section \ref{subsec: CoR non-naive}.
\end{proposition}

The EIFs in Proposition \ref{prop: EIF for CoR} slightly generalize those in \citet{branson2023causal}. Cross-fitted, one-step estimators of $\tau_{\omega_{s}^{\text{s-trim}}}^{\text{num}}(a, s)$ and $\tau_{\omega_{s}^{\text{s-trim}}}^{\text{den}}(a, s)$ can be constructed using EIFs in Proposition \ref{prop: EIF for CoR} as follows:
$$
\widehat{\tau}_{\omega_{s}^{\text{s-trim}}}^{\text{num}}(a, s) = \frac{1}{n} \sum_{k=1}^{K} \sum_{i \in \mathcal{I}_{n,k}} \widehat{\varphi}^{\text{num}}(a, s),
\text { and  } 
\widehat{\tau}_{\omega_{s}^{\text{s-trim}}}^{\text{den}}(a, s) = \frac{1}{n} \sum_{k=1}^{K} \sum_{i \in \mathcal{I}_{n,k}} \widehat{\varphi}^{\text{den}}(a, s),
$$
where $\{\mathcal{I}_{n,k}\}_{k = 1, \ldots, K}$ of $\{1, \ldots, n\}$ is a $K$-fold random partition for some fixed $K$. For each $k \in \{1,2, \ldots, K\}$, samples in $\mathcal{I}^{\complement}_{n,k} = \{1, \ldots, n\} \backslash \mathcal{I}_{n,k}$ constitute the training set, and the nuisance functions in $\widehat{\varphi}^{\text{num}}(a, s)$ and $\widehat{\varphi}^{\text{den}}(a, s)$ are estimated via parametric, semiparametric, or flexible machine learning methods fitted on the training set; see Supplemental Material \ref{app:crossfitting-tau} for details. 

Finally, an estimator of $\text{STWCR}(a, s)$, denoted $\widehat{\tau}(a, s)$, can be obtained as the ratio of $\widehat{\tau}_{\omega_{s}^{\text{s-trim}}}^{\text{num}}(a, s)$ and $\widehat{\tau}_{\omega_{s}^{\text{s-trim}}}^{\text{den}}(a, s)$ as follows: 
$$
\widehat{\tau}(a, s)=\widehat{\tau}_{\omega_{s}^{\text{s-trim}}}^{\text{num}}(a, s) / \widehat{\tau}_{\omega_{s}^{\text{s-trim}}}^{\text{den}}(a, s).
$$


Proposition \ref{prop: asymp-CoR} establishes the properties of the proposed estimator $\widehat{\tau}(a, s)$.

\begin{proposition}
\label{prop: asymp-CoR}
Under regularity conditions in Supplemental Material \ref{app: regularity condition for prop cor} and for fixed $t$, $h$, and $\epsilon$, we have
\[
\sqrt{n}\Bigl(
\widehat{\tau}_{\omega_s^{\mathrm{s\text{-}trim}}}^{\mathrm{num}}(a,s)
-
\tau_{\omega_s^{\mathrm{s\text{-}trim}}}^{\mathrm{num}}(a,s)
\Bigr)
\overset{d}{\longrightarrow}
N\!\Bigl(0,\,
\mathrm{Var}\{\varphi^{\mathrm{num}}(a,s)\}
\Bigr),
\]
\[
\sqrt{n}\Bigl(
\widehat{\tau}_{\omega_s^{\mathrm{s\text{-}trim}}}^{\mathrm{den}}(a,s)
-
\tau_{\omega_s^{\mathrm{s\text{-}trim}}}^{\mathrm{den}}(a,s)
\Bigr)
\overset{d}{\longrightarrow}
N\!\Bigl(0,\,
\mathrm{Var}\{\varphi^{\mathrm{den}}(a,s)\}
\Bigr).
\]
Moreover, we have
\[
\sqrt{n}\{\widehat{\tau}(a,s) - \mathrm{STWCR}(a,s)\}
\overset{d}{\longrightarrow}
N\!\left(0,\,
\sigma_{1}^2
\right),
\]
where
\[
\sigma^2_1
=
\mathrm{Var}\!\left(
\frac{
\varphi^{\mathrm{num}}(a,s)
-
\tau(a,s)\,
\varphi^{\mathrm{den}}(a,s)
}{
\tau^{\mathrm{den}}(a,s)
}
\right).
\]
The variance $\sigma^2_1$ can be consistently estimated via a plug-in estimator. 
\end{proposition}

Propositions~\ref{prop: EIF for CoR} and \ref{prop: asymp-CoR} establish the EIFs and associated EIF-based estimators for $\mathrm{STWCR}(a, s)$, which serve as the building blocks for $\mathrm{STWCRVE}(a_1, a_0, s_1, s_0)$ that contrasts two STWCR parameters. 

For fixed $t$, $h_0$, $h_1$, and $\epsilon$, 
Theorem \ref{thm: EIF CVE} derives the uncentered EIFs for $\tau_{\omega_{s_1, s_0}^{\text{sd-trim}}}^{\text{num}}(a_1, s_1)$ and $\tau_{\omega_{s_1, s_0}^{\text{sd-trim}}}^{\text{den}}(a_0, s_0)$ in $\mathrm{STWCRVE}(a_1, a_0, s_1, s_0)$.

\begin{theorem}
\label{thm: EIF CVE}
The uncentered efficient influence functions for $\tau_{\omega_{s_1, s_0}^{\text{sd-trim}}}^{\text{num}}(a_1, s_1)$ and $\tau_{\omega_{s_1, s_0}^{\text{sd-trim}}}^{\text{den}}(a_0, s_0)$ are
{\scriptsize
\begin{align*}
    &\theta^{num}(a_1, s_1) \\
    &= \mathbb{IF}\left(\tau_{\omega_{s_1, s_0}^{\text{sd-trim}}}^{\text{num}}(a_1, s_1)\right) + \tau_{\omega_{s_1, s_0}^{\text{sd-trim}}}^{\text{num}}(a_1, s_1)\\
    &= \int_{s^{\prime}\in \mathcal{S}} K_{h_0}(s' - s_0) K_{h_1}(S - s_1)
\frac{I(A = a_1)}{\pi'(a_1 \mid B, \bs{X})} \frac{\partial \phi\left\{\pi(S \mid a_1, B, \bs{X}), t\right\}}{\partial \pi(S \mid a_1, B, \bs{X})} \phi\left\{\pi(s' \mid a_0, B, \bs{X}), t\right\} \, r(a_1, S, B, \bs{X}) \, ds' \\[8pt]
&\quad + \int_{s^{\prime\prime}\in \mathcal{S}} K_{h_0}(S - s_0) K_{h_1}(s'' - s_1)
\frac{I(A = a_0)}{\pi'(a_0 \mid B, \bs{X})}
\frac{\partial \phi\left\{\pi(S \mid a_0, B, \bs{X}), t\right\}}{\partial \pi(S \mid a_0, B, \bs{X})}
\phi\left\{\pi(s'' \mid a_1, B, \bs{X}), t\right\} \, r(a_1, s'', B, \bs{X}) \, ds'' \\[8pt]
&\quad +
\int_{s^{\prime\prime}\in \mathcal{S}} K_{h_0}(S - s_0) K_{h_1}(s'' - s_1)
\frac{I(A = a_1)}{\pi'(a_1 \mid B, \bs{X}) \, \pi(S \mid a_1, B, \bs{X})}
\phi\left\{\pi(S \mid a_0, B, \bs{X}), t\right\} \phi\left\{\pi(S \mid a_1, B, \bs{X}), t\right\}
\big(Y - r(a_1, S, B, \bs{X})\big) \, ds'' \\[8pt]
&\quad -
\iint_{s^{\prime}, s^{\prime\prime}\in \mathcal{S}} K_{h_0}(s' - s_0) K_{h_1}(s'' - s_1)
\Bigg[
\frac{I(A = a_1)}{\pi'(a_1 \mid B, \bs{X})}
\frac{\partial \phi\left\{\pi(s'' \mid a_1, B, \bs{X}), t\right\}}{\partial \pi}
\pi(s'' \mid a_1, B, \bs{X})
\phi\left\{\pi(s' \mid a_0, B, \bs{X}), t\right\}
r(a_1, s'', B, \bs{X}) \\[4pt]
&\hspace{4cm} +
\frac{I(A = a_0)}{\pi'(a_0 \mid B, \bs{X})}
\frac{\partial \phi\left\{\pi(s' \mid a_0, B, \bs{X}), t\right\}}{\partial \pi}
\pi(s' \mid a_0, B, \bs{X})
\phi\left\{\pi(s'' \mid a_1, B, \bs{X}), t\right\}
r(a_1, s'', B, \bs{X})
\Bigg] ds' ds'' \\[8pt]
&\quad +
\iint_{s^{\prime}, s^{\prime\prime}\in \mathcal{S}} K_{h_0}(s' - s_0) K_{h_1}(s'' - s_1)
\phi\left\{\pi(s' \mid a_0, B, \bs{X}), t\right\}
\phi\left\{\pi(s'' \mid a_1, B, \bs{X}), t\right\}
r(a_1, s'', B, \bs{X}) \, ds' ds'', \\
\end{align*}}
and
{\scriptsize
\begin{align*}
    &\theta^{den}(a_0, s_0)\\
    &=\mathbb{IF}\left(\tau_{\omega_{s_1, s_0}^{\text{sd-trim}}}^{\text{den}}(a_0, s_0)\right) + \tau_{\omega_{s_1, s_0}^{\text{sd-trim}}}^{\text{den}}(a_0, s_0)\\
    &=\int_{s^{\prime}\in \mathcal{S}} K_{h_0}(s' - s_0) K_{h_1}(S - s_1)
\frac{I(A = a_1)}{\pi'(a_1 \mid B, \bs{X})}
\frac{\partial \phi\left\{\pi(S \mid a_1, B, \bs{X}), t\right\}}{\partial \pi(S \mid a_1, B, \bs{X})}
\phi\left\{\pi(s' \mid a_0, B, \bs{X}), t\right\} \, r(a_0, s', B, \bs{X}) \, ds' \\[8pt]
&\quad +
\int_{s^{\prime\prime}\in \mathcal{S}} K_{h_0}(S - s_0) K_{h_1}(s'' - s_1)
\frac{I(A = a_0)}{\pi'(a_0 \mid B, \bs{X})}
\frac{\partial \phi\left\{\pi(S \mid a_0, B, \bs{X}), t\right\}}{\partial \pi(S \mid a_0, B, \bs{X})}
\phi\left\{\pi(s'' \mid a_1, B, \bs{X}), t\right\} \, r(a_0, S, B, \bs{X}) \, ds'' \\[8pt]
&\quad +
\int_{s^{\prime\prime}\in \mathcal{S}} K_{h_0}(S - s_0) K_{h_1}(s'' - s_1)
\frac{I(A = a_0)}{\pi'(a_0 \mid B, \bs{X}) \, \pi(S \mid a_0, B, \bs{X})}
\phi\left\{\pi(S \mid a_0, B, \bs{X}), t\right\} \phi\left\{\pi(s'' \mid a_1, B, \bs{X}), t\right\}
\big(Y - r(a_0, S, B, \bs{X})\big) \, ds'' \\[8pt]
&\quad -
\iint_{s^{\prime}, s^{\prime\prime}\in \mathcal{S}} K_{h_0}(s' - s_0) K_{h_1}(s'' - s_1)
\Bigg[
\frac{I(A = a_1)}{\pi'(a_1 \mid B, \bs{X})}
\frac{\partial \phi\left\{\pi(s'' \mid a_1, B, \bs{X}), t\right\}}{\partial \pi}
\pi(s'' \mid a_1, B, \bs{X})
\phi\left\{\pi(s' \mid a_0, B, \bs{X}), t\right\}
r(a_0, s', B, \bs{X}) \\[4pt]
&\hspace{4cm} +
\frac{I(A = a_0)}{\pi'(a_0 \mid B, \bs{X})}
\frac{\partial \phi\left\{\pi(s' \mid a_0, B, \bs{X}), t\right\}}{\partial \pi}
\pi(s' \mid a_0, B, \bs{X})
\phi\left\{\pi(s'' \mid a_1, B, \bs{X}), t\right\}
r(a_0, s', B, \bs{X})
\Bigg] ds' ds'' \\[8pt]
&\quad +
\iint_{s^{\prime}, s^{\prime\prime}\in \mathcal{S}} K_{h_0}(s' - s_0) K_{h_1}(s'' - s_1)
\phi\left\{\pi(s' \mid a_0, B, \bs{X}), t\right\}
\phi\left\{\pi(s'' \mid a_1, B, \bs{X}), t\right\}
r(a_0, s', B, \bs{X}) \, ds' ds'',
\end{align*}}
respectively, where $\pi'(a \mid B, \bs{X})$, $\pi(S \mid a, B, \bs{X})$, and $\phi\{\cdot, t\}$ are defined as in Proposition \ref{prop: EIF for CoR}, and $K_{h_0}(\cdot)$ and $K_{h_1}(\cdot)$ are kernel functions with bandwidtch $h_0$ and $h_1$, respectively.
\end{theorem}

The uncertered efficient influence functions $\theta^{\text{num}}(a_1, s_1)$ and $\theta^{\text{den}}(a_0, s_0)$ can then be used to construct the estimator for the target parameter $\text{STWCRVE}(a_1, a_0, s_1, s_0)$ as follows:
\begin{align*}
    \widehat{\delta}(a_1, a_0, s_1, s_0) :=  1 - \frac{\widehat{\tau}_{\omega_{s_1, s_0}^{\text{sd-trim}}}^{\text{num}}(a_1, s_1)}{\widehat{\tau}_{\omega_{s_1, s_0}^{\text{sd-trim}}}^{\text{den}}(a_0, s_0)},
\end{align*}
where
$$\widehat{\tau}_{\omega_{s_1, s_0}^{\text{sd-trim}}}^{\text{num}}(a_1, s_1) = \frac{1}{n} \sum_{k=1}^{K} \sum_{i \in \mathcal{I}_{n,k}}\widehat{\theta}^{\text{num}}(a_1, s_1) \text { and } \widehat{\tau}_{\omega_{s_1, s_0}^{\text{sd-trim}}}^{\text{den}}(a_0, s_0) = \frac{1}{n} \sum_{k=1}^{K} \sum_{i \in \mathcal{I}_{n,k}}\widehat{\theta}^{\text{den}}(a_0, s_0)
$$
are cross-fitted one-step estimators of $\theta^{\text{num}}(a_1, s_1)$ and $\theta^{\text{den}}(a_0, s_0)$, respectively.

Theorem \ref{thm: EIF_CVE_property} establishes the theoretical properties of the proposed estimator $\widehat{\delta}(a_1, a_0, s_1, s_0)$.

\begin{theorem}
\label{thm: EIF_CVE_property}
Under regularity conditions in Supplemental Material \ref{app: regularity condition for thm cve} and for fixed $t$, $h_0$, $h_1$, and $\epsilon$, we have
\[
\sqrt{n}\bigl\{\widehat{\delta}(a_1, a_0, s_1,s_0)-\mathrm{STWCRVE}(a_1, a_0, s_1,s_0)\bigr\}
\overset{d}{\longrightarrow}
N\!\left(0,\sigma^2_2\right),
\]
where
\[
\sigma^2_2
=
\mathrm{Var}\!\left(
\frac{
\theta^{\mathrm{num}}(a_1,s_1)
-
\delta(a_1, a_0, s_1,s_0)\,\theta^{\mathrm{den}}(a_0,s_0)
}{
\theta^{\mathrm{den}}(a_0,s_0)
}
\right).
\]
The variance $\sigma^2_2$ can be consistently estimated via a plug-in estimator, and a $100\times (1-\alpha)\%$ confidence interval for 
$\mathrm{STWCRVE}(a_1, a_0, s_1,s_0)$ can be constructed accordingly.

Alternatively, note that when
$
\rho(a_1,a_0,s_1,s_0)
=
\frac{
\tau_{\omega_{s_1, s_0}^{\text{sd-trim}}}^{\text{num}}(a_1, s_1)
}{
\tau_{\omega_{s_1, s_0}^{\text{sd-trim}}}^{\text{den}}(a_0, s_0)
}
>0$, inference can be conducted on the log scale. By the delta method,
\[
\sqrt{n}\left\{
\log
\frac{
\widehat{\tau}_{\omega_{s_1, s_0}^{\text{sd-trim}}}^{\text{num}}(a_1, s_1)
}{
\widehat{\tau}_{\omega_{s_1, s_0}^{\text{sd-trim}}}^{\text{den}}(a_0, s_0)
}
-
\log
\frac{
\tau_{\omega_{s_1, s_0}^{\text{sd-trim}}}^{\text{num}}(a_1, s_1)
}{
\tau_{\omega_{s_1, s_0}^{\text{sd-trim}}}^{\text{den}}(a_0, s_0)
}
\right\}
\overset{d}{\longrightarrow}
N\!\left(0,\sigma_{2,\log}^2\right),
\]
where
\[
\sigma_{2,\log}^2
=
\mathrm{Var}\!\left(
\frac{\theta^{\mathrm{num}}(a_1,s_1)}
{\tau_{\omega_{s_1, s_0}^{\text{sd-trim}}}^{\text{num}}(a_1, s_1)}
-
\frac{\theta^{\mathrm{den}}(a_0,s_0)}
{\tau_{\omega_{s_1, s_0}^{\text{sd-trim}}}^{\text{den}}(a_0, s_0)}
\right).
\]
Let $\widehat{\sigma}_{2,\log}$ denote a plug-in estimator of $\sigma_{2,\log}$, and define $\widehat{\rho}(a_1,a_0,s_1,s_0)
=
\frac{
\widehat{\tau}_{\omega_{s_1, s_0}^{\text{sd-trim}}}^{\text{num}}(a_1, s_1)
}{
\widehat{\tau}_{\omega_{s_1, s_0}^{\text{sd-trim}}}^{\text{den}}(a_0, s_0)
}.$
Then a $100\times(1-\alpha)\%$ confidence interval for $\rho(a_1,a_0,s_1,s_0)$ is
\[[L_{\rho},U_{\rho}] = 
\left[
\widehat{\rho}(a_1,a_0,s_1,s_0)
\exp\!\left\{
- z_{1-\alpha/2}\frac{\widehat{\sigma}_{2,\log}}{\sqrt{n}}
\right\},
\;
\widehat{\rho}(a_1,a_0,s_1,s_0)
\exp\!\left\{
z_{1-\alpha/2}\frac{\widehat{\sigma}_{2,\log}}{\sqrt{n}}
\right\}
\right],
\]
and the corresponding
$100\times(1-\alpha)\%$ confidence interval for
$\mathrm{STWCRVE}(a_1,a_0,s_1,s_0)$ is
$[\,1-U_{\rho},\ 1-L_{\rho}\,].$
\end{theorem}

%% file: 4.Simulation.tex
\section{Simulation}
\label{sec: Simu}
\subsection{Simulation settings}
\label{subsec: simu setting}
We generate data for a hypothetical non-na\"ive study population modeled after the COVAIL study population. The first factor we vary is the sample size:
\begin{description}
    \item[Factor 1:] Number of study participants, $n$: $1000$, $2000$, and $5000$.
\end{description}
For each study participant, we generate three covariates $\bs{X} = (X_1, X_2, X_3)$, where $X_1 \sim \text{Bernoulli}(0.3)$ represents a participant's prior exposure status ($X_1 = 0$ for na\"ive and $X_1 = 1$ for non-na\"ive), and $X_2, X_3 \sim \text{Uniform}[0,1]$ correspond to two baseline risk factors. 

In addition, we generate a baseline immune marker $B$ for each participant. The second factor we vary is the distribution of $B$ (\textbf{Factor 2}): 
\begin{description}
        \item[Scenario I: Discrete $B$.] The baseline immune marker $B$ is sampled from a categorical distribution that takes values $\{1, 2, 3, 4, 5\}$ with probability $p = (0.2, 0.3, 0.4, 0.05, 0.05)$ among na\"ive participants  and with probability $p = (0.1, 0.15, 0.3, 0.3, 0.15)$ among non-na\"ive participants.

        \item[Scenario II: Continuous $B$.] The baseline immune marker $B$ follows a Gamma distribution with shape parameter $2.5$ and rate parameter $1.0$ among na\"ive participants. Among non-na\"ive participants, $B$ follows a Gamma distribution with shape parameter $3.0$ and rate parameter $0.7$. To avoid extreme values, the top $0.5\%$ of $B$ values are truncated at its $99.5$th percentile.
        
        \item[Scenario III: Discrete $B$ with a large probability mass at 0 for na\"ive participants.] The baseline immune marker $B$ is sampled from a categorical distribution that takes values $\{0, 1, 2, 3, 4\}$ with probability $p = (0.6, 0.2, 0.1, 0.05, 0.05)$ among na\"ive participants  and with probability $p = (0.1, 0.15, 0.3, 0.3, 0.15)$ among non-na\"ive participants.
\end{description}
Each participant is randomized to an investigational vaccine $(A = 1)$ or a comparator vaccine $(A = 0)$. The peak immune marker $S$ is generated from the following structural model:
\[
S = B + A - 0.5X_1 + X_2^2 + 4 + \varepsilon, \quad \varepsilon \sim \mathcal{N}(0,1).
\]
According to this model, most participants would have $S > B.$ Participants who receive the investigational vaccine ($A = 1$) tend to have a slightly larger boost in their peak immune response level compared to those who receive the comparator vaccine ($A = 0$).  Figure \ref{fig: simu S vs B} in the Supplemental Material \ref{app: add simu rslts} plots the joint distribution of $(B, S)$ for participants with $X_2 = 0.5$.

Finally, we generate a binary infectious disease outcome defined by the presence of qualifying symptoms together with molecular confirmation of the causative pathogen; for simplicity, we hereafter refer to this outcome as ``infection.” The clinical outcome $Y$ is generated as follows:
\[
\text{logit}\{P(Y=1)\} = 0.5X_2 +2 X_3 - 0.2S - A - 0.3B + 1.5,
\]
where $\text{logit}(p) = \log\{p/(1-p)\}$. According to this model, participants with a higher baseline immune marker $B$, receiving an investigational vaccine, and a higher peak immune marker, tend to have a lower probability of infection.

For each simulated dataset, with $t = 0.1$, $h = 0.1$, and $\epsilon = 0.1$ fixed, we calculated the proposed EIF-based estimator $\widehat{\tau}(1, s)$ of the smoothed trimmed weighted controlled risk, $\text{STWCR(1, s)}$, for selected values of $s$, and constructed the corresponding 95\% confidence interval based on the asymptotic distribution in Proposition \ref{prop: asymp-CoR}. By setting $t=0.1$, the estimand evaluates the controlled risk among individuals in the overall population who had at least a 10\% probability of achieving the post-vaccination marker level of interest. We then calculated the proposed EIF-based estimator $\widehat{\delta}(a_1, a_0, s_1, s_0)$ of the smoothed trimmed weighted controlled relative vaccine efficacy, $\text{STWCRVE}(a_1, a_0, s_1, s_0)$, for selected values of $(s_1, s_0)$ and with $t = 0.1$, $h_1 = 0.1$, $h_0 = 0.1$, and $\epsilon = 0.1$ fixed. The 95\% confidence intervals of $\text{STWCRVE}(a_1, a_0, s_1, s_0)$ were constructed using the log-transformation as described in Theorem \ref{thm: EIF_CVE_property}.

For each of the $3 \times 3 = 9$ simulation settings defined by sample size (\textbf{Factor 1}) and distribution of $B$ (\textbf{Factor 2}), we repeated the simulation $1000$ times. The performance of the proposed estimators was evaluated based on the percentage of bias and empirical coverage of the $95\%$ confidence intervals.

\subsection{Simulation results}
\label{subsec: simu CoR}

\begin{figure}[ht]
    \centering
    \includegraphics[width=\linewidth]{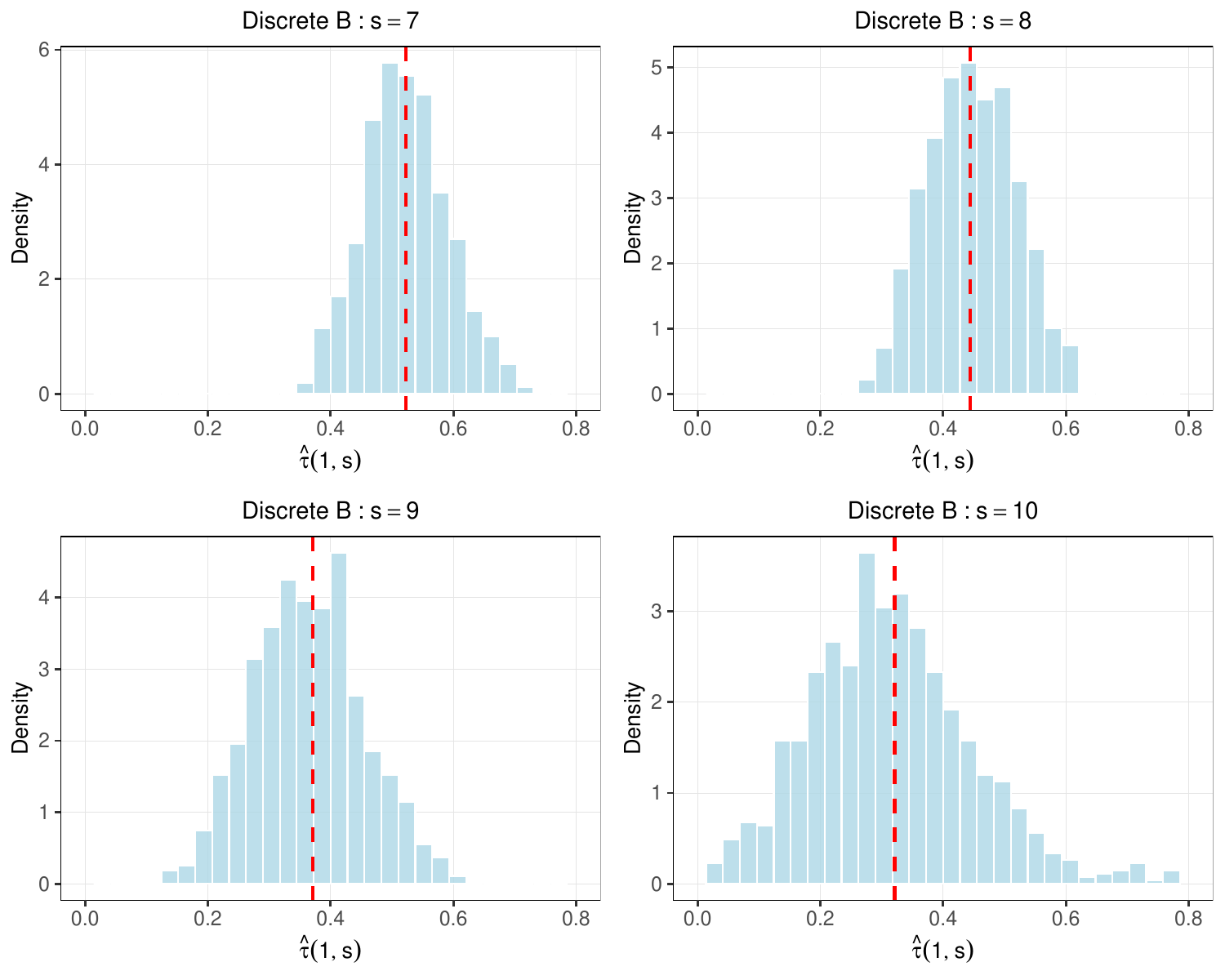}
    \caption{Sampling distributions of EIF-based estimator $\widehat{\tau}(1, s)$ of the smoothed trimmed weighted controlled risk, $\text{STWCR(1, s)}$, for selected values of $s$ when B is discrete. The sample size is $1000$ and simulations were repeated $1000$ times. The dashed red lines represent the ground truth in each case.}
    \label{fig:EIF distribution B discrete}
\end{figure}

Figure \ref{fig:EIF distribution B discrete} displays the sampling distribution of the proposed estimator $\widehat{\tau}(1,s)$ evaluated at $s = 7$, $s = 8$, $s = 9$, and $s = 10$, when the sample size $n = 1000$ and the baseline immune marker $B$ is discrete (\textsf{Scenario I}). The ground truth value of $\text{STWCR}(1, s)$ was superimposed as a red dashed line. The figure confirms that the proposed estimator is approximately normally distributed. Figure \ref{fig:EIF distribution B cont} in the Supplemental Material \ref{app: add simu rslts} plots the analogous figure when $n = 1000$ and $B$ is continuous.

Panel A of Table \ref{tbl:sim_bias_cover} summarizes the percentage bias and empirical coverage of the 95\% confidence intervals across various simulation settings and different choices of $s$ for $\widehat{\tau}(1,s)$. Two consistent patterns emerge. First, the proposed estimator exhibits little bias and has coverage close to the nominal level across most simulation settings. Second, for a fixed sample size (e.g., $n=2000$), the estimator performs worse when $s$ is near the boundary. For example, under \textsf{Scenario I} when $n=2000$, the empirical coverage of the confidence intervals is close to the nominal level for $s=7$ and $s=8$, but declines to 93.7\% at $s=9$ and further to 92.6\% at $s=10$. This pattern is expected because, for boundary values (e.g., $s=10$), the relevant subpopulation, those with at least a 10\% probability of achieving $s$ as large as $10$ after receiving $A = 1$, is small, so most of the population does not contribute to estimation or inference. When $s$ is near the boundary (e.g., $s = 9$ or $10$), increasing the sample size, for instance, from $2000$ to $5000$, appears to help improve the performance of the proposed estimator.

\begin{table}[ht]
\centering
\scriptsize
\resizebox{0.99\textwidth}{!}{
\begin{tabular}{ccccccccccccc}
  \hline \\[-8pt]
   &\multicolumn{6}{c}{\textbf{Panel A}: STWCR $\widehat{\tau}(1, s)$ }
    &\multicolumn{6}{c}{\textbf{Panel B}: STWCRVE $\widehat{\delta}(a_1, a_0, s_1, s_0)$} \\ \cline{2-13}
    &\multicolumn{2}{c}{\textsf{Scenario I}}
    &\multicolumn{2}{c}{\textsf{Scenario II}}
    &\multicolumn{2}{c}{\textsf{Scenario III}}
    &\multicolumn{2}{c}{\textsf{Scenario I}}
    &\multicolumn{2}{c}{\textsf{Scenario II}}
    &\multicolumn{2}{c}{\textsf{Scenario III}}\\  \cline{2-13}
    \multirow{3}{*}{\begin{tabular}{c} \textsf{Sample} \\ \textsf{size} \end{tabular}}
    &\multirow{3}{*}{\begin{tabular}{c} \textsf{Perc.}\\ \textsf{bias}\end{tabular}}
    &\multirow{3}{*}{\begin{tabular}{c} \textsf{Cov.}\end{tabular}}
    &\multirow{3}{*}{\begin{tabular}{c} \textsf{Perc.}\\ \textsf{bias}\end{tabular}}
     &\multirow{3}{*}{\begin{tabular}{c} \textsf{Cov.} \end{tabular}}
    &\multirow{3}{*}{\begin{tabular}{c} \textsf{Perc.}\\ \textsf{bias}\end{tabular}}
     &\multirow{3}{*}{\begin{tabular}{c} \textsf{Cov.} \end{tabular}}
    &\multirow{3}{*}{\begin{tabular}{c} \textsf{Perc.}\\ \textsf{bias}\end{tabular}}
    &\multirow{3}{*}{\begin{tabular}{c} \textsf{Cov.} \end{tabular}}
    &\multirow{3}{*}{\begin{tabular}{c} \textsf{Perc.}\\ \textsf{bias}\end{tabular}}
    &\multirow{3}{*}{\begin{tabular}{c} \textsf{Cov.} \end{tabular}}
    &\multirow{3}{*}{\begin{tabular}{c} \textsf{Perc.}\\ \textsf{bias}\end{tabular}}
    &\multirow{3}{*}{\begin{tabular}{c} \textsf{Cov.} \end{tabular}}\\ \\ \\
  \hline \\[-8pt]
&\multicolumn{6}{c}{$s = 7$} & \multicolumn{6}{c}{$(s_0,s_1) = (7,8)$}\\[2pt]
1000 & -0.1\% & 94.5\% &  0.3\% & 94.7\% & -0.21\% & 93.9\% & -1.42\% & 96.3\% & -1.14\% & 96.1\% & -2.9\% & 95.9\%\\
2000 &  0.6\% & 95.0\% &  1.0\% & 95.3\% &  0.13\% & 94.5\% & -1.46\% & 94.8\% & -2.31\% & 96.7\% & -1.16\% & 96.3\%\\
5000 & -0.02\% & 95.7\% &  0.5\% & 95.1\% &  0.51\% & 94.5\% & -0.22\% & 94.4\% &  0.10\% & 95.9\% & -0.13\% & 95.0\%\\[2pt]

&\multicolumn{6}{c}{$s = 8$} &\multicolumn{6}{c}{$(s_0,s_1) = (7,10)$}\\[2pt]
1000 &  0.6\% & 94.9\% & -0.4\% & 94.2\% &  0.68\% & 93.5\% & -1.83\% & 90.7\% & -2.32\% & 91.8\% &  1.32\% & 90.0\%\\
2000 &  0.2\% & 95.4\% & -1.3\% & 94.3\% & -1.31\% & 93.8\% &  0.16\% & 93.4\% & -2.44\% & 92.4\% & -1.36\% & 89.8\%\\
5000 & -0.2\% & 94.3\% & -0.8\% & 94.5\% & -0.81\% & 95.5\% & -1.20\% & 96.1\% &  0.54\% & 95.5\% &  0.99\% & 94.1\%\\[2pt]

&\multicolumn{6}{c}{$s = 9$} &\multicolumn{6}{c}{$(s_0,s_1) = (8,8)$}\\
1000 & -2.2\% & 93.7\% & -3.4\% & 91.1\% & -5.96\% & 92.8\% & -1.72\% & 95.5\% & -2.08\% & 96.3\% & -1.05\% & 96.2\%\\
2000 & -0.9\% & 93.7\% & -2.2\% & 94.7\% & -0.17\% & 94.2\% & -3.23\% & 95.1\% & -4.15\% & 95.8\% & -0.19\% & 95.0\%\\
5000 & -0.5\% & 96.2\% & -1.6\% & 95.5\% & -0.80\% & 96.6\% & -0.89\% & 96.0\% & -1.82\% & 95.6\% & -1.07\% & 96.1\%\\[2pt]

&\multicolumn{6}{c}{$s = 10$} &\multicolumn{6}{c}{$(s_0,s_1) = (8,10)$}\\
1000 &  3.0\% & 88.5\% & -7.8\% & 87.0\% & -4.97\% & 85.0\% & -2.34\% & 93.9\% & -1.08\% & 94.8\% &  2.34\% & 91.7\%\\
2000 &  0.6\% & 92.6\% & -7.9\% & 89.5\% & -3.52\% & 90.5\% & -0.53\% & 94.6\% & -2.72\% & 94.5\% & -0.25\% & 93.7\%\\
5000 &  0.9\% & 95.1\% & -8.7\% & 94.3\% & -4.40\% & 92.8\% & -1.23\% & 96.1\% & -0.33\% & 96.1\% & -0.03\% & 95.0\%\\
\hline
\end{tabular}}
\caption{Percentage bias (\textsf{Perc. bias}) and empirical coverage of $95\%$ confidence intervals (\textsf{Cov.}) of the proposed estimators $\widehat{\tau}(1, s)$ for STWCR across various choices of $s$ (\textbf{Panel A}) and $\widehat{\delta}(a_1, a_0, s_1, s_0)$ for STWCRVE across various combinations of $(s_1, s_0)$ (\textbf{Panel B}) under different simulation settings. \textsf{Scenario I}: discrete $B$. \textsf{Scenario II}: continuous $B$. \textsf{Scenario III}: discrete $B$ with a large probability mass at 0 for na\"ive participants.}
\label{tbl:sim_bias_cover}
\end{table}

Panel B of Table \ref{tbl:sim_bias_cover} further summarizes the performance of $\widehat{\delta}(a_1, a_0, s_1, s_0)$ across multiple choices of $(s_1, s_0)$. Overall, the estimator $\widehat{\delta}(a_1, a_0, s_1, s_0)$ demonstrates performance qualitatively similar to that of $\widehat{\tau}(1, s)$. For a fixed sample size, the finite-sample performance of the estimator tends to deteriorate as either $s_1$ or $s_0$ approaches boundary values. For example, under \textsf{Scenario I} when $n = 1000$ and $s_0 = 7$, the empirical coverage of the 95\% confidence intervals declines from $96.3\%$ to $90.7\%$ as $s_1$ increases from 8 to 10. For fixed values of $(s_1, s_0)$, the percentage bias generally decreases and the empirical coverage improves as the sample size increases. For example, under \textsf{Scenario I} with $(s_0, s_1) = (7, 10)$, the coverage increases from $90.7\%$ to $96.1\%$ as $n$ grows from $1000$ to $5000$.

%% file: 5.Case_study.tex
\section{Coronavirus Variant Immunologic Landscape (COVAIL) trial}
\label{sec: case study}
\subsection{Background: trial schema, objectives, and immune correlates of risk}
\label{subsec: case study background}
The Coronavirus Variant Immunologic Landscape (COVAIL) trial enrolled approximately 1250 adults in the United States who had completed a primary COVID-19 vaccination series and received one prior booster dose. Participants received a second booster between March~30 and October~28,~2022, across four sequential enrollment stages. Across these stages, participants were randomized to one of 17 booster vaccine arms: six Moderna mRNA vaccine arms in Stage~1 (approximately 100 participants per arm), six Pfizer--BioNTech mRNA vaccine arms in Stage~2 (approximately 50 participants per arm), three Sanofi recombinant protein vaccine arms in Stage~3 (approximately 50 participants per arm), and two Pfizer--BioNTech bivalent mRNA vaccine arms in Stage~4 (approximately 100 participants per arm). Vaccines differed by the number of SARS-CoV-2 strains included in the construct (monovalent versus bivalent) and by the lineage of the encoded strains.

The primary objective of the COVAIL trial was to evaluate immunogenicity across booster vaccines with differing variant formulations. The collection of clinical COVID-19 endpoints additionally enabled the identification of immune correlates of risk and protection following booster vaccination. Using pseudovirus neutralizing antibody ID\textsubscript{50} titers (nAb-ID\textsubscript{50}) measured against multiple SARS-CoV-2 variants, \citet{zhang2025neutralizing} demonstrated a consistent inverse association between neutralizing antibody levels measured at a prespecified, near-peak timepoint (15~days post-vaccination) and the risk of COVID-19 over approximately six months of follow-up. The strength of this association varied by prior infection status, with stronger correlates observed among previously infected participants. A similar association was observed among Stage~3 participants who received the Sanofi booster \citep{fong2025neutralizing}.

\subsection{Correlation between D1 and D15 nAb-ID\textsubscript{50} titers}
\label{subsec: case study descriptives}
\begin{figure}[ht]
  \centering
  \includegraphics[width=\textwidth]{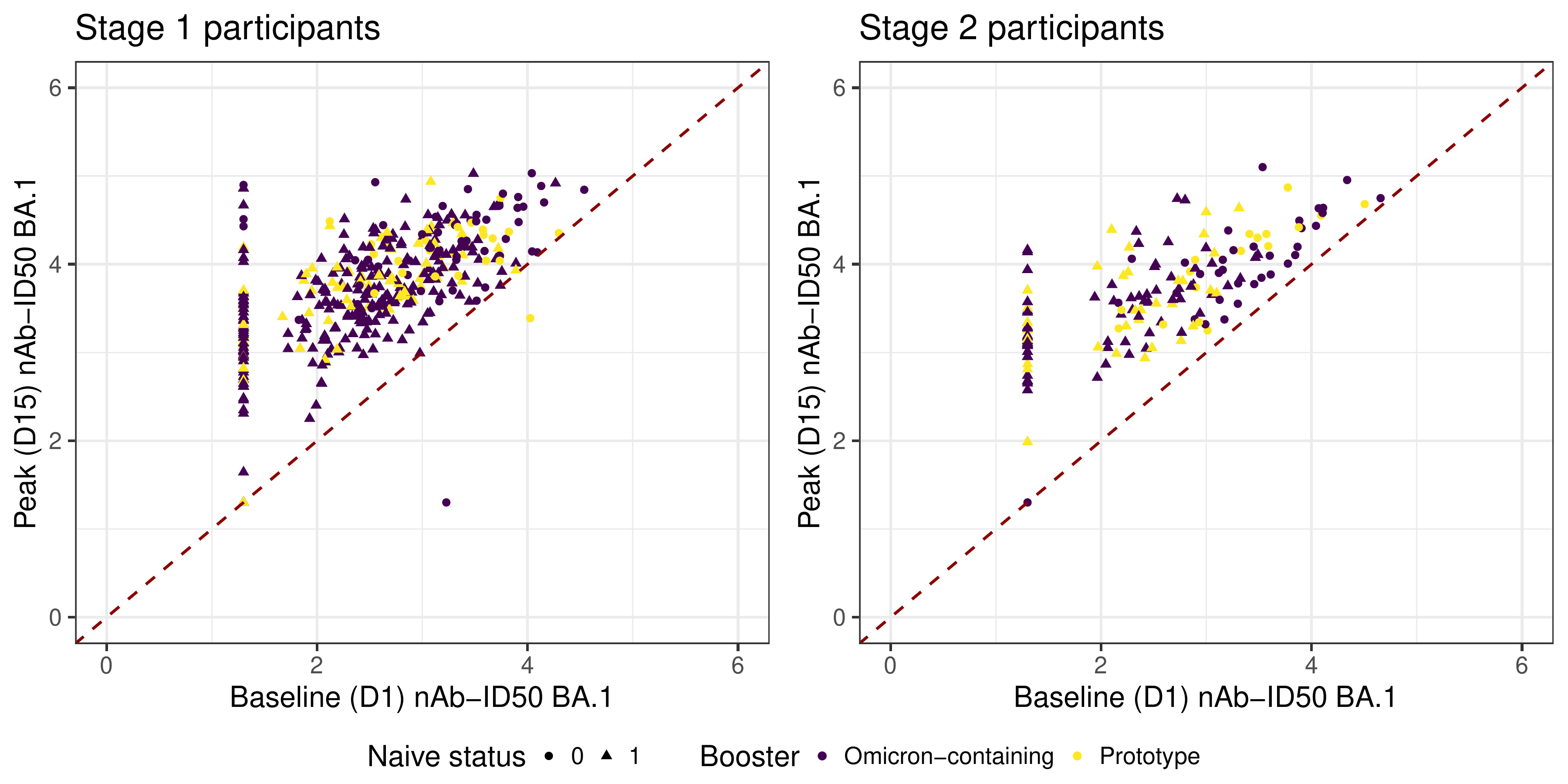}
  \caption{The joint distribution of baseline (Day 1, D1) and approximately peak (Day 15, D15) nAb-ID\textsubscript{50} titers against Omicron BA.1 among Stage 1 (left panel) and Stage 2 (right panel) participants.}
  \label{fig: case study S vs B}
\end{figure}

Figure~\ref{fig: case study S vs B} shows the joint distribution of baseline (Day~1, or D1, prior to study vaccine administration) and peak (Day~15, or D15, measured 15 days post--study vaccine) nAb-ID\textsubscript{50} levels against Omicron BA.1 among participants in Stage~1 (left panel) and Stage~2 (right panel). The analysis is restricted to per-protocol participants in the COVAIL trial who received a single-dose mRNA booster vaccination, following \citet{zhang2025neutralizing}. As expected, nearly all observations lie above the diagonal dashed line, indicating that peak (D15) nAb-ID\textsubscript{50} levels generally exceeded baseline (D1) levels. For subsequent analyses, we excluded two participants whose D15 levels were substantially lower than their D1 levels, as these measurements were suspected to be erroneous readouts from the assay.

\subsection{Controlled risk CoP analysis among Stage-1 participants}
\label{subsec: case study controlled risk}
\begin{figure}[ht]
  \centering
  \includegraphics[width=\textwidth]{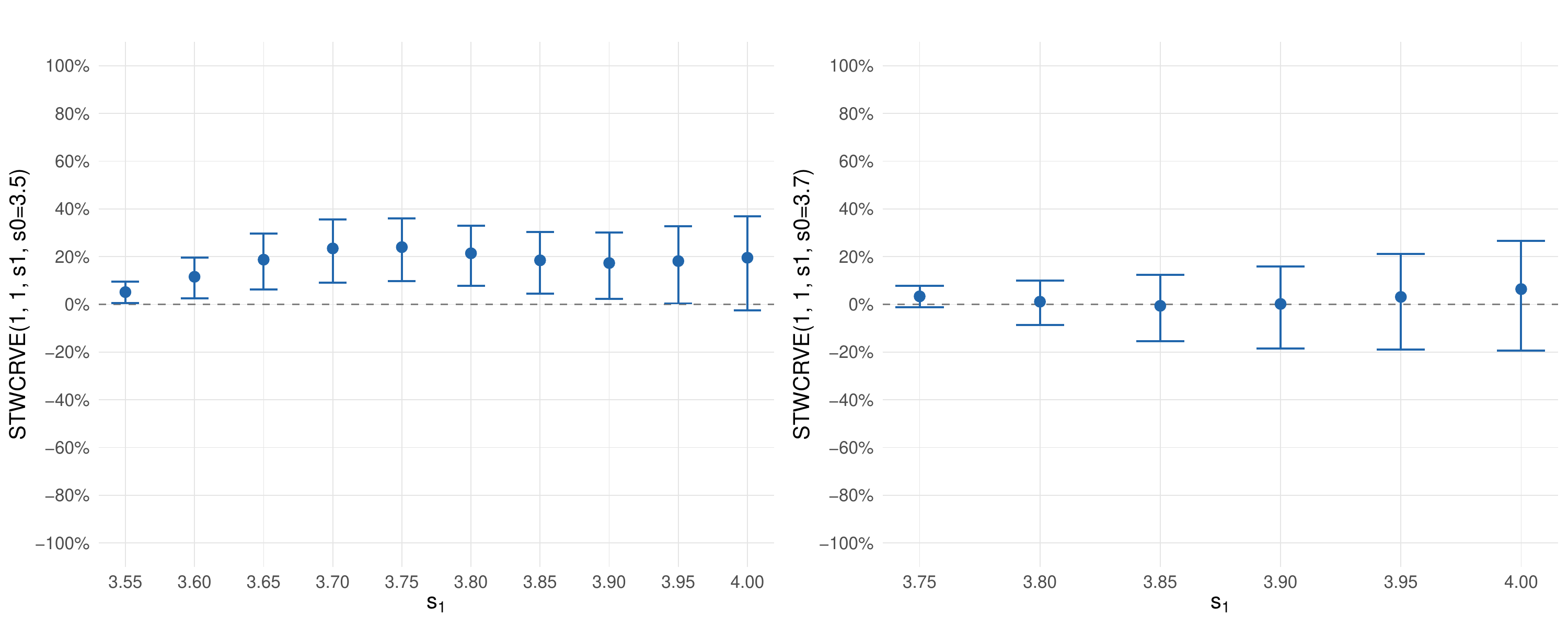}
  \caption{Estimated $\mathrm{STWCRVE}(1, 1, s_1, s_0 = 3.5)$ (left panel) and $\mathrm{STWCRVE}(1, 1, s_1, s_0 = 3.7)$ (right panel) with pointwise 95\% confidence intervals for selected values of $s_1 > s_0$. }
  \label{fig: case study controlled risk CoP}
\end{figure}

We evaluated peak nAb-ID\textsubscript{50} as a controlled risk CoP among Stage-1 participants who received an Omicron-containing vaccine. The left panel of Figure~\ref{fig: case study controlled risk CoP} displays the estimated $\mathrm{STWCRVE}(1, 1, s_1, s_0 = 3.5)$ for selected values of $s_1 > 3.5$. For example, when $s_1 = 3.8$, among the subpopulation with at least a $10\%$ probability of achieving either a peak nAb-ID\textsubscript{50} of $3.5$ or $3.8$, the controlled risk was lower when the peak titer was set to $3.8$ compared with $3.5$, resulting in a positive relative risk. The remaining confidence intervals in the left panel of Figure \ref{fig: case study controlled risk CoP} and those in the right panel can be interpreted analogously. Overall, our findings indicate that higher peak nAb-ID\textsubscript{50} titers, relative to lower titers, result in lower controlled risk within the same target subpopulation.

\subsection{Controlled direct effects analysis}
\label{subsec: case study controlled direct effect}
\begin{figure}[ht]
  \centering
  \includegraphics[width=0.65\textwidth]{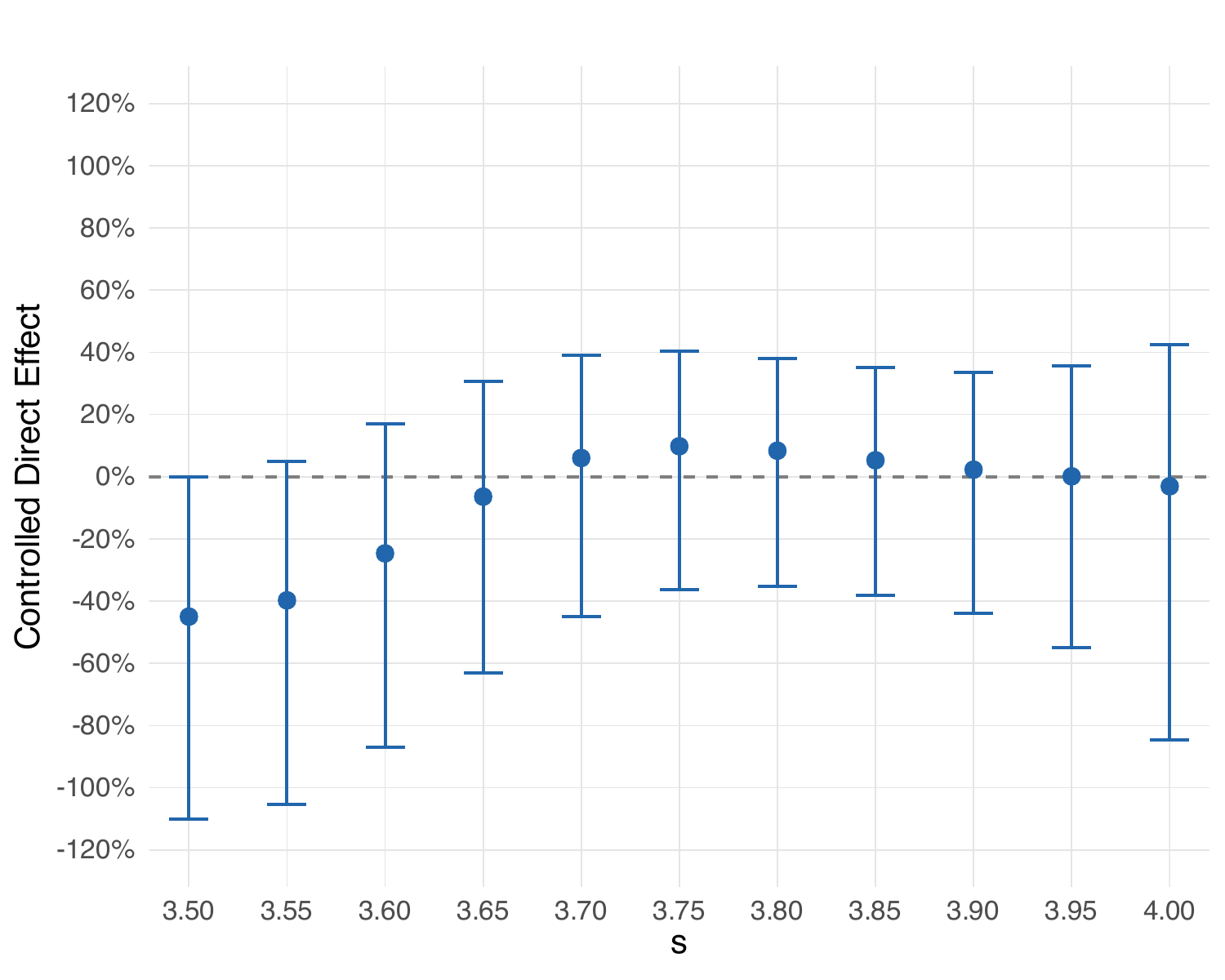}
  \caption{Controlled direct effect with pointwise 95\% confidence intervals comparing Omicron-containing to Prototype vaccine recipients among participants who have an at least 10\% probability of attaining D15 nAb-ID\textsubscript{50} level of $s$. }
  \label{fig: controlled direct effect}
\end{figure}

Figure~\ref{fig: controlled direct effect} plots the estimated controlled direct effects comparing Omicron-containing vaccines with the Prototype vaccine among Stage-1 participants who have at least a 10\% probability of attaining D15 nAb-ID\textsubscript{50} level $S = s$ under either vaccine across a range of $s$ values. These controlled direct effects were obtained by setting $s_1 = s_0 = s$ and letting $t = 0.1$ in the smoothed trimmed weighted controlled relative vaccine efficacy, $\mathrm{STWCRVE}(1, 0, s_1, s_0)$. Figure \ref{fig: controlled direct effect} suggests no evidence of controlled direct effects across various Stage-1 subpopulations, each corresponding to a different D15 nAb-ID\textsubscript{50} level. Notably, confidence intervals are generally wider for values near the extremes of the spectrum (e.g., $s = 3.5$ or $s = 4$), which reflects the smaller sample sizes of the subpopulations associated with these values of $s$. 

Finally, Table \ref{tab:cve_s0_s1} in the Supplemental Material \ref{app: case study} summarizes estimated $\mathrm{STWCRVE}(1, 0, s_1, s_0)$ for selected $(s_1, s_0)$ values when $s_1 \neq s_0$. For instance, among Stage-1 participants who had an at least 10\% probability of achieving a $3.75$-log D15 nAb-ID\textsubscript{50} titer when receiving the Omicron-containing vaccine and at least 10\% probability of achieving a $3.5$-log D15 nAb-ID\textsubscript{50} titer when receiving the Prototype vaccine, the controlled relative VE was estimated to be $24.2\%$ (95\%: -28.0\% to 55.2\%; subpopulation controlled risk = 33.1\% vs. 43.7\%). 

%% file: 6.Discussion.tex
\section{Discussion}
\label{sec: discussion}
Understanding how the infectious disease risk would vary with the vaccine-elicited immune marker level is a key question in immune correlates research. In many recent immune correlates work whose study populations consisted of participants with varying levels of baseline immunity, it is a conundrum whether or not to adjust for baseline immune marker levels. Not adjusting for them could potentially leave out an important confounder of the mediator-outcome relationship: baseline immunity is often highly associated with the post-vaccination immune marker level and is also likely to be associated with the disease endpoint risk. On the other hand, conditioning on the baseline immune response level makes the positivity assumption dubious, because the post-vaccination immune response level is highly unlikely to fall below the baseline level. 

We propose one approach to address this conundrum. The method builds upon a controlled effects approach to immune correlates analysis. The method acknowledges that the potential outcome $Y(A = a, S = s)$ is only well defined for participants who have a positive probability of achieving post-vaccination immune response level $S = s$ after receiving vaccine $A = a$, and use a data-adaptive weighting scheme to focus on these participants when estimating the potential outcomes, or contrast of two potential outcomes. 

The techniques used in the paper were adapted from the literature on propensity score trimming in a non-mediation, point exposure context \citep{crump2009dealing,yang2018asymptotic,branson2023causal}. While the violation of the positivity assumption is often discussed in a point exposure setting, it has received relatively little attention in mediation analysis. In many biomedical studies, the distribution of the mediator would depend on baseline characteristics of the study participants in a unidirectional way, and similar violation of the positivity assumption is likely to arise. Apart from the proposed method, stochastic-intervention-based methods are a promising alternative, although these methods seek to answer a different question \citep{hejazi2021efficient,huang2023stochastic}. Other ways to deal with the positivity violation are worth exploring.

One important limitation of propensity score--trimming--based methods, including the proposed adaptation to mediation analysis in this article, is that the target study population depends on the specified level of the continuous treatment in a non-mediation, point-exposure setting, or on the mediator level in the present context. \citet{branson2023causal} showed that, for each treatment level, the corresponding subpopulation can be characterized in the sense that any moment of the baseline covariates within that subpopulation is identifiable. Nevertheless, such subpopulations are often not directly ``actionable'' in clinical practice, because researchers cannot determine whether a given participant belongs to the subpopulation associated with a particular treatment or mediator level. Despite this limitation, estimating the subpopulation-specific relative risk,
$\mathrm{STWCRVE}(a, a, s_1, s_0),$ can still provide valuable insight into the role of immune markers in reducing disease risk and into underlying immune mechanisms, as illustrated in our case study of the COVAIL trial.

%% file: Appendix.tex



\clearpage

\setcounter{page}{1}

\pagestyle{plain}

\setcounter{section}{0}
\setcounter{subsection}{0}
\setcounter{equation}{0}
\setcounter{figure}{0}
\setcounter{table}{0}

\renewcommand{\thesection}{S\arabic{section}}
\renewcommand{\thesubsection}{S\arabic{section}.\arabic{subsection}}
\renewcommand{\theequation}{S\arabic{equation}}
\renewcommand{\thefigure}{S\arabic{figure}}
\renewcommand{\thetable}{S\arabic{table}}

\begin{center}
{\Large \bf Supplemental Materials to ``Dealing with positivity violations in mediation analysis via weighted controlled effects, with application to assessing immune correlates of protection in antigen-experienced participants" by He and Zhang. }\\[0.5em]
\end{center}

\vspace{1em}

\section{Proofs}

\subsection{Proof of Proposition \ref{prop: EIF for CoR}}

Let $\mathcal{X}'$ denote the support of $\bs{X}'=(B,\bs{X})$, and let $\mathcal{S}\subseteq\mathbb{R}$ denote the support of $S$. 
Let $\bs{X}^\prime = (B, \bs{X})$. We have the following influence functions:

$$
\begin{aligned}
\mathbb{I F}\{r(a, s, x^\prime)\} & =\frac{\mathbb{I}(\bs{X}^\prime=x^\prime, S=s, A=a)}{\pi(s \mid a, x^\prime) \cdot \pi^\prime(a \mid x^\prime) \cdot p(x^\prime)}(Y-r(a, s, x^\prime)), \\
\mathbb{I F}\{p(x^\prime)\} & =\mathbb{I}(\bs{X}^\prime=x^\prime)-p(x^\prime),\\
\mathbb{I F}\{\pi(s \mid a, x^\prime)\} & =\frac{\mathbb{I}(\bs{X}^\prime = x^\prime, A=a)}{P(\bs{X}^\prime = x^\prime, A=a)}\left(\mathbb{I}(S=s)-\pi(s \mid a, x^\prime )\right),\\
\mathbb{I F}\{\pi^\prime(a \mid x^\prime)\} & =\frac{\mathbb{I}(\bs{X}^\prime = x^\prime)}{p(x^\prime)}\left(\mathbb{I}(A=a)-\pi^\prime(a \mid x^\prime)\right).\\
\mathbb{IF}\left(S(\pi(s \mid a, x^\prime), t)\right) 
&= \frac{\partial S(\pi(s \mid s, x^\prime), t)}{\partial \pi} \mathbb{IF}\{\pi(s \mid a, x^\prime)\}\\
&= \frac{\partial S(\pi(s \mid a, x^\prime), t)}{\partial \pi} \frac{ I(\bs{X}^\prime = x^\prime, A=a)}{P(\bs{X}^\prime = x^\prime, A=a)}\left[I(S=s) - \pi(s \mid a, x^\prime)\right].\\
\end{aligned}
$$

Then,

\begin{align*}
\mathbb{I F}(\tau_{\omega_{s}^{\text{s-trim}}}^{\text{den}}(a, s))
= & \mathbb{I F}\left[\sum_{x^{\prime}\in \mathcal{X}'} \sum_{s_0\in \mathcal{S}} K_h\left(s_0-s\right) S\left(\pi\left(s_0 \mid a, x^{\prime}\right), t\right) p\left(x^{\prime}\right)\right] \\
= & \sum_{x^{\prime}\in \mathcal{X}'} \sum_{s_0\in \mathcal{S}} K_h\left(s_0-s\right) \mathbb{I F}\left(S\left(\pi\left(s_0 \mid a, x^{\prime}\right), t\right) p\left(x^{\prime}\right)\right) \\
= & \sum_{x^{\prime}\in \mathcal{X}'} \sum_{s_0\in \mathcal{S}} K_h\left(s_0 -s\right)\mathbb{I F}\left(S\left(\pi\left(s_0 \mid a, x^{\prime}\right), t\right) \right) p\left(x^{\prime}\right) \\
&+ \sum_{x^{\prime}\in \mathcal{X}'} \sum_{s_0\in \mathcal{S}} K_h\left(s_0-s\right) S\left(\pi\left(s_0 \mid a, x^{\prime}\right), t\right)  \mathbb{I F}\left(p\left(x^{\prime}\right)\right) \\
= & \sum_{x^{\prime}\in \mathcal{X}'} \sum_{s_0\in \mathcal{S}} K_h\left(s_0-s\right) \frac{\partial S\left(\pi\left(s_0 \mid a, x^{\prime}\right), t\right)}{\partial \pi} \cdot 
\frac{I\left(\bs{X}^\prime=x^{\prime}, A=a\right)}{p\left(\bs{X}^\prime=x^{\prime}, A=a\right)}
\left(I\left(S=s_0\right)-\pi\left(s_0 \mid a, x^{\prime}\right)\right) \\
& \cdot p(x^\prime) \\
&+\sum_{x^{\prime}\in \mathcal{X}'} \sum_{s_0\in \mathcal{S}} K_h\left(s_0-s\right) S\left(\pi\left(s_0 \mid a, x^{\prime}\right), t\right) \cdot \left[I\left(\bs{X}^\prime=x^{\prime}\right)-p\left(x^{\prime}\right)\right]\\
=&\sum_{x^\prime\in \mathcal{X}'} \sum_{s_0\in \mathcal{S}} K_h\left(s_0-s\right) 
\frac{\partial S\left(\pi\left(s_0 \mid a, x^\prime\right), t\right)}{\partial \pi\left(s_0 \mid a, x^{\prime}\right)} 
\frac{I\left(\bs{X}^\prime=x^{\prime}, A=a\right)}{\pi^\prime(a|x^\prime)}
\left[I\left(S=s_0\right)-\pi\left(s_0 \mid a, x^{\prime}\right)\right] \\
& +\sum_{x^\prime\in \mathcal{X}'} \sum_{s_0\in \mathcal{S}} K_h\left(s_0-s\right) 
S\left(\pi\left(s_0 \mid a, x^{\prime}\right), t\right)
\left[I(\bs{X}^\prime=x^\prime)-p\left(x^{\prime}\right)\right]\\
=&K_h(S-s) \frac{I(A = a)}{\pi^\prime(a|\bs{X})}
\frac{\partial S\left(\pi\left(S \mid a, \bs{X}^\prime\right), t\right)}{\partial \pi\left(S \mid a, \bs{X}^\prime\right)}\\
& - \int_{s_0\in \mathcal{S}} K_h\left(s_0-s\right) \frac{I(A = a)}{\pi^\prime(a|\bs{X}^\prime)} 
\frac{\partial S\left(\pi\left(s_0 \mid a, \bs{X}^\prime\right), t\right)}{\partial \pi\left(s_0 \mid a, \bs{X}^\prime\right)} 
\pi\left(s_0 \mid a, \bs{X}^\prime\right)\, d s_0 \\
& +\int_{s_0\in \mathcal{S}} K_h\left(s-s_0\right) S\left(\pi\left(s_0 \mid a, \bs{X}^{\prime}\right), t\right)\, {d s_0}
-\tau_{\omega_{s}^{\text{s-trim}}}^{\text{den}}(a, s) \\
\end{align*}

and

$$
\begin{aligned}
\mathbb{I F}(\tau_{\omega_{s}^{\text{s-trim}}}^{\text{num}}(a, s))
&=\mathbb{I F}\left(\sum_{x^{\prime}\in\mathcal{X}'} \sum_{s_0\in\mathcal{S}} K_h\left(s_0-s\right) 
S\left(\pi\left(s_0 \mid a, x^{\prime}\right), t\right) r\left(a, s_0, x^{\prime}\right) p\left(x^{\prime}\right)\right) \\
=&\sum_{x^{\prime}\in\mathcal{X}'} \sum_{s_0\in\mathcal{S}} K_h\left(s_0-s\right)\, \mathbb{I F}(S)\cdot r\cdot p \quad(1) \\
&+\sum_{x^{\prime}\in\mathcal{X}'} \sum_{s_0\in\mathcal{S}} K_h\left(s_0-s\right)\, \mathbb{I F}(r)\cdot S\cdot p \quad(2)\\
&+\sum_{x^{\prime}\in\mathcal{X}'} \sum_{s_0\in\mathcal{S}} K_h\left(s_0-s\right)\, \mathbb{I F}(p)\cdot S\cdot r \quad(3).
\end{aligned}
$$
where,

$$
\begin{aligned}
(1)
&=\sum_{x^{\prime}\in\mathcal{X}'} \sum_{s_0\in\mathcal{S}} K_h\left(s_0-s\right) 
\frac{\partial S\left(\pi\left(s_0 \mid a, x^{\prime}\right), t\right)}{\partial \pi\left(s_0 \mid a, x^{\prime}\right)} 
\frac{I\left(\bs{X}^{\prime}=x^{\prime}, A=a\right)}{\pi^{\prime}\left(a \mid x^{\prime}\right)}
\left[I\left(S=s_0\right)-\pi\left(s_0 \mid a, x^{\prime}\right)\right] \cdot r\left(a,s_0,x'\right) \\
& =K_h(S - s) \frac{I(A=a)}{\pi^{\prime}\left(a \mid \bs{X}^{\prime}\right)} 
\frac{\partial S\left(\pi\left(S \mid a, \bs{X}^{\prime}\right), t\right)}{\partial \pi\left(S \mid a, \bs{X}^{\prime}\right)} 
r\left(a, S, \bs{X}^{\prime}\right) \\
&\quad -\int_{s_0\in\mathcal{S}} K_h\left(s_0-s\right) \frac{I(A = a)}{\pi^{\prime}\left(a \mid \bs{X}^{\prime}\right)} 
\frac{\partial S\left(\pi\left(s_0 \mid a, \bs{X}^{\prime}\right), t\right)}{\partial \pi\left(s_0 \mid a, \bs{X}^{\prime}\right)} 
\pi\left(s_0 \mid a, \bs{X}^{\prime}\right) r\left(a, s_0, \bs{X}^{\prime}\right)\, d s_0 .
\end{aligned}
$$

$$
\begin{aligned}
(2)
&=\sum_{x^{\prime}\in\mathcal{X}'} \sum_{s_0\in\mathcal{S}} K_h\left(s_0- s\right) 
\frac{I\left(\bs{X}^{\prime}=x^{\prime}, S=s_0, A=a\right)}
{\pi\left(s_0 \mid a, x^{\prime}\right)\pi^{\prime}\left(a \mid x^{\prime}\right)}
\left(Y-r\left(a, s_0, x^{\prime}\right)\right) \cdot S\left(\pi(s_0\mid a,x'),t\right)\\
&=K_h\left(S-s\right)\frac{I(A=a)}{\pi^{\prime}\left(a \mid \bs{X}^{\prime}\right)}
\frac{ S\left(\pi\left(S \mid a, \bs{X}^{\prime}\right), t\right)}{ \pi\left(S \mid a, \bs{X}^{\prime}\right)} 
\left(Y - r(a, S, \bs{X}^\prime)\right).
\end{aligned}
$$

$$
\begin{aligned}
(3) 
& =\sum_{x^{\prime}\in\mathcal{X}'} \sum_{s_0\in\mathcal{S}} K_h\left(s_0-s\right) 
S\left(\pi\left(s_0 \mid a, x^{\prime}\right), t\right)\cdot \left[I\left(\bs{X}^{\prime}=x^{\prime}\right)-p\left(x^{\prime}\right)\right]\cdot r\left(a,s_0,x'\right) \\
& =\int_{s_0\in\mathcal{S}} K_h\left(s-s_0\right) 
S\left(\pi\left(s_0 \mid a, \bs{X}^{\prime}\right), t\right) 
r\left(a, s_0, \bs{X}^{\prime}\right)\, d s_0 
-\tau_{\omega_{s}^{\text{s-trim}}}^{\text{num}}(a, s).
\end{aligned}
$$
Thus, 
$$
\begin{aligned}
\mathbb{I F}(\tau_{\omega_{s}^{\text{s-trim}}}^{\text{num}}(a, s))
&=K_h(S - s) \frac{I(A=a)}{\pi^{\prime}\left(a \mid \bs{X}^{\prime}\right)} 
\frac{\partial S\left(\pi\left(S \mid a, \bs{X}^{\prime}\right), t\right)}{\partial \pi\left(S \mid a, \bs{X}^{\prime}\right)} 
r\left(a, S, \bs{X}^{\prime}\right) \\
&\quad +K_h\left(S-s\right)\frac{I(A=a)}{\pi^{\prime}\left(a \mid \bs{X}^{\prime}\right)}
\frac{ S\left(\pi\left(S \mid a, \bs{X}^{\prime}\right), t\right)}{ \pi\left(S \mid a, \bs{X}^{\prime}\right)} 
\left(Y - r(a, S, \bs{X}^\prime)\right)\\
&\quad -\int_{s_0\in\mathcal{S}} K_h\left(s_0-s\right) \frac{I(A = a)}{\pi^{\prime}\left(a \mid \bs{X}^{\prime}\right)} 
\frac{\partial S\left(\pi\left(s_0 \mid a, \bs{X}^{\prime}\right), t\right)}{\partial \pi\left(s_0 \mid a, \bs{X}^{\prime}\right)} 
\pi\left(s_0 \mid a, \bs{X}^{\prime}\right) r\left(a, s_0, \bs{X}^{\prime}\right)\, d s_0\\
&\quad +\int_{s_0\in\mathcal{S}} K_h\left(s-s_0\right) 
S\left(\pi\left(s_0 \mid a, \bs{X}^{\prime}\right), t\right) 
r\left(a, s_0, \bs{X}^{\prime}\right)\, d s_0 \\
&\quad -\tau_{\omega_{s}^{\text{s-trim}}}^{\text{num}}(a, s).
\end{aligned}
$$

\subsection{Proof of Theorem \ref{thm: EIF CVE}}
We have
$$
\begin{aligned}
 \tau^{\text{num}}_{\omega_{s_1, s_0}^{\text{sd-trim}}}(a_1, s_1) = &
\iiiint_{\substack{s^\prime\in\mathcal{S},\, s^{\prime\prime}\in\mathcal{S},\\ \bs{X}^\prime\in\mathcal{X}'}}
K_{h_0}(s^\prime - s_0)\,K_{h_1}(s^{\prime \prime} - s_1)\cdot 
S({\pi}(s^{\prime \prime} \mid a_1, B, \bs{X}), t)\,
S({\pi}(s^\prime \mid a_0, B, \bs{X}), t) \\
&\cdot r(a_1, s^{\prime \prime}, B, \bs{X})\,ds^\prime\, ds^{\prime \prime}\, dP(B)\, dP(\bs{X}),
\end{aligned}
$$
and
$$
\begin{aligned}
\tau^{\text{den}}_{\omega_{s_1, s_0}^{\text{sd-trim}}}(a_0, s_0) = &
\iiiint_{\substack{s^\prime\in\mathcal{S},\, s^{\prime\prime}\in\mathcal{S},\\ \bs{X}^\prime\in\mathcal{X}'}}
K_{h_0}(s^\prime - s_0)\,K_{h_1}(s^{\prime \prime} - s_1)\cdot 
S({\pi}(s^{\prime \prime} \mid a_1, B, \bs{X}), t)\,
S({\pi}(s^\prime \mid a_0, B, \bs{X}), t) \\
&\cdot r(a_0, s^\prime, B, \bs{X})\,ds^\prime\, ds^{\prime \prime}\, dP(B)\, dP(\bs{X}).
\end{aligned}
$$
Then, 

\begin{align*}
\mathbb{I F}(\tau^{\text{den}}_{\omega_{s_1, s_0}^{\text{sd-trim}}}(a_0, s_0))
    =&\mathbb{I F}\left(\sum_{s^\prime\in\mathcal{S}} \sum_{s^{\prime \prime}\in\mathcal{S}} \sum_{x^\prime\in\mathcal{X}'} 
\left[K_{h_0}(s^\prime - s_0)K_{h_1}(s^{\prime \prime} - s_1)\right. \right. \\
&\left. \left. \cdot S({\pi}(s^{\prime \prime} \mid a_1,  x^\prime), t) 
S({\pi}(s^\prime \mid a_0,  x^\prime), t) 
\cdot r(a_0, s^\prime,  x^\prime)p(x^\prime)\right]\right)\\
=&\sum_{s^\prime\in\mathcal{S}} \sum_{s^{\prime \prime}\in\mathcal{S}} \sum_{x^\prime\in\mathcal{X}'} 
K_{h_0}(s^\prime - s_0)K_{h_1}(s^{\prime \prime} - s_1)\mathbb{I F}\left[\right.  \\
&\left. \cdot S({\pi}(s^{\prime \prime} \mid a_1,  x^\prime), t) 
S({\pi}(s^\prime \mid a_0,  x^\prime), t) 
\cdot r(a_0, s^\prime,  x^\prime)p(x^\prime)\right].\\
\end{align*}
Generally,
\begin{align*}
&\mathbb{I F}\left[S({\pi}(s^{\prime \prime} \mid a_1,  x^\prime), t) 
S({\pi}(s^{\prime} \mid a_0,  x^\prime), t) \cdot r(a_0, s^\prime,  x^\prime)p(x^\prime)\right],
\qquad 
s',s''\in\mathcal{S},\; x'\in\mathcal{X}'\\
&= \mathbb{I F}\left[S({\pi}(s^{\prime \prime} \mid a_1,  x^\prime), t)\right]
S({\pi}(s^{\prime} \mid a_0,  x^\prime), t) \cdot r(a_0, s^\prime,  x^\prime)p(x^\prime) \quad (1)\\
&\quad + \mathbb{I F}\left[S({\pi}(s^{\prime} \mid a_0,  x^\prime), t)\right]
S({\pi}(s^{\prime \prime} \mid a_1,  x^\prime), t) \cdot r(a_0, s^\prime,  x^\prime)p(x^\prime) \quad (2)\\
&\quad +S({\pi}(s^{\prime} \mid a_0,  x^\prime), t)
S({\pi}(s^{\prime \prime} \mid a_1,  x^\prime), t) \cdot  
\mathbb{I F}\left[r(a_0, s^\prime,  x^\prime)\right]p(x^\prime) \quad (3)\\
&\quad +S({\pi}(s^{\prime} \mid a_0,  x^\prime), t)
S({\pi}(s^{\prime \prime} \mid a_1,  x^\prime), t) \cdot 
r(a_0, s^\prime,  x^\prime) \mathbb{I F}\left[p(x^\prime)\right] \quad (4).
\end{align*}
where
\begin{align*}
    (1)& = \frac{\partial S({\pi}(s^{\prime \prime} \mid a_1, x^\prime), t)}{\partial {\pi}(s^{\prime \prime} \mid a_1, x^\prime)} \frac{ I(\bs{X}^\prime = x^\prime, A=a_1)}{P(\bs{X}^\prime = x^\prime, A=a_1)}\left[I(S=s^{\prime \prime}) - \pi(s^{\prime \prime} \mid a_1, x^\prime)\right]S({\pi}(s^{\prime} \mid a_0,  x^\prime), t) r(a_0, s^\prime,  x^\prime)p(x^\prime)\\
    &= \frac{\partial S({\pi}(s^{\prime \prime} \mid a_1, x^\prime), t)}{\partial \pi} \frac{ I(\bs{X}^\prime = x^\prime, A=a_1)}{{\pi^\prime}(a_1 \mid  x^\prime)}\left[I(S=s^{\prime \prime}) - \pi(s^{\prime \prime} \mid a_1, x^\prime)\right]S({\pi}(s^{\prime} \mid a_0,  x^\prime), t) r(a_0, s^\prime,  x^\prime),
\end{align*}
\begin{align*}
    (2)& =\frac{\partial S({\pi}(s^{\prime} \mid a_0, x^\prime), t)}{\partial \pi} \frac{ I(\bs{X}^\prime = x^\prime, A=a_0)}{P(\bs{X}^\prime = x^\prime, A=a_0)}\left[I(S=s^\prime) - \pi(s^\prime \mid a_0, x^\prime)\right]S({\pi}(s^{\prime \prime} \mid a_1,  x^\prime), t) r(a_0, s^\prime,  x^\prime)p(x^\prime)\\
    &= \frac{\partial S({\pi}(s^{\prime} \mid a_1, x^\prime), t)}{\partial \pi} \frac{ I(\bs{X}^\prime = x^\prime, A=a_0)}{{\pi^\prime}(a_0 \mid  x^\prime)}\left[I(S=s^{\prime}) - \pi(s^{\prime} \mid a_0, x^\prime)\right]S({\pi}(s^{\prime \prime} \mid a_1,  x^\prime), t) r(a_1, s^\prime,  x^\prime),
\end{align*}
\begin{align*}
    (3) &= \frac{\mathbb{I}(\bs{X}^\prime=x^\prime, S=s^\prime, A=a_0)}{\pi(s^\prime \mid a_0, x^\prime) \pi^\prime(a_0 \mid x^\prime) \cdot p(x^\prime)}(Y-r(a_0, s^\prime, x^\prime)) S({\pi}(s^{\prime} \mid a_0,  x^\prime), t)S({\pi}(s^{\prime \prime} \mid a_1,  x^\prime), t) p(x^\prime)\\
    & = \frac{\mathbb{I}(\bs{X}^\prime=x^\prime, S=s^\prime, A=a_0)}{\pi(s^\prime \mid a_0, x^\prime){\pi^\prime}(a_0 \mid  x^\prime)}(Y-r(a_0, s^\prime, x^\prime)) S({\pi}(s^{\prime} \mid a_0,  x^\prime), t)S({\pi}(s^{\prime \prime} \mid a_1,  x^\prime), t),
\end{align*}
and
\begin{align*}
    (4) &=\left[\mathbb{I}(\bs{X}^\prime=x^\prime)-p(x^\prime)\right]S({\pi}(s^{\prime} \mid a_0,  x^\prime), t)S({\pi}(s^{\prime \prime} \mid a_1,  x^\prime), t) \cdot r(a_0, s^\prime,  x^\prime).\\
\end{align*}
Then,
Let $\bs{X}^\prime=(B,\bs{X})$ have support $\mathcal{X}'$, and let $S$ have support $\mathcal{S}\subseteq\mathbb{R}$.

\begin{align*}
&\sum_{s^\prime\in\mathcal{S}} \sum_{s^{\prime \prime}\in\mathcal{S}} \sum_{x^\prime\in\mathcal{X}'} 
K_{h_0}(s^\prime - s_0)K_{h_1}(s^{\prime \prime} - s_1) \big[(1) + (2)\big] \\
&=\sum_{s^\prime\in\mathcal{S}} \sum_{s^{\prime \prime}\in\mathcal{S}} \sum_{x^\prime\in\mathcal{X}'} 
K_{h_0}(s^\prime - s_0)K_{h_1}(s^{\prime \prime} - s_1) 
\frac{\partial S({\pi}(s^{\prime \prime} \mid a_1, x^\prime), t)}{\partial \pi} 
\frac{ I(\bs{X}^\prime = x^\prime, A=a_1)}{{\pi^\prime}(a_1 \mid  x^\prime)}
\left[I(S=s^{\prime \prime}) - \pi(s^{\prime \prime} \mid a_1, x^\prime)\right]\\
&\quad \cdot S({\pi}(s^{\prime} \mid a_0,  x^\prime), t)\, r(a_0, s^\prime,  x^\prime)\\
&\quad + \sum_{s^\prime\in\mathcal{S}} \sum_{s^{\prime \prime}\in\mathcal{S}} \sum_{x^\prime\in\mathcal{X}'} 
K_{h_0}(s^\prime - s_0)K_{h_1}(s^{\prime \prime} - s_1) 
\frac{\partial S({\pi}(s^{\prime} \mid a_0, x^\prime), t)}{\partial \pi} 
\frac{ I(\bs{X}^\prime = x^\prime, A=a_0)}{{\pi^\prime}(a_0 \mid  x^\prime)}
\left[I(S=s^{\prime}) - \pi(s^{\prime} \mid a_0, x^\prime)\right]\\
&\quad \cdot S({\pi}(s^{\prime \prime} \mid a_1,  x^\prime), t)\, r(a_0, s^\prime,  x^\prime)\\
&=\int_{s^\prime\in\mathcal{S}} K_{h_0}(s^\prime - s_0)K_{h_1}(S- s_1) 
\frac{\partial S({\pi}(S \mid a_1, \bs{X}^\prime), t)}{\partial \pi} 
\frac{ I(A=a_1)}{\pi^\prime(a_1 \mid  \bs{X}^\prime)}
S({\pi}(s^{\prime} \mid a_0,  \bs{X}^\prime), t)\, r(a_0, s^\prime,  \bs{X}^\prime)\, ds^\prime\\
&-\iint_{\substack{s^\prime\in\mathcal{S},\, s^{\prime\prime}\in\mathcal{S}}}
K_{h_0}(s^\prime - s_0)K_{h_1}(s^{\prime \prime} - s_1) 
\frac{\partial S({\pi}(s^{\prime \prime} \mid a_1, \bs{X}^\prime), t)}{\partial \pi} 
\frac{ I(A=a_1)}{\pi^\prime(a_1 \mid  \bs{X}^\prime)}
\pi(s^{\prime \prime} \mid a_1, \bs{X}^\prime)
S({\pi}(s^{\prime} \mid a_0,  \bs{X}^\prime), t)\,\\
& \cdot r(a_0, s^\prime,  \bs{X}^\prime)\, ds^\prime ds^{\prime \prime}\\
&+\int_{s^{\prime\prime}\in\mathcal{S}} K_{h_0}(S - s_0)K_{h_1}(s^{\prime \prime}- s_1) 
\frac{\partial S({\pi}(S \mid a_0, \bs{X}^\prime), t)}{\partial \pi} 
\frac{ I(A=a_0)}{\pi^\prime(a_0 \mid  \bs{X}^\prime)}
S({\pi}(s^{\prime \prime} \mid a_1,  \bs{X}^\prime), t)\, r(a_0, S,  \bs{X}^\prime)\, ds^{\prime \prime}\\
&-\iint_{\substack{s^\prime\in\mathcal{S},\, s^{\prime\prime}\in\mathcal{S}}}
K_{h_0}(s^\prime - s_0)K_{h_1}(s^{\prime \prime} - s_1) 
\frac{\partial S({\pi}(s^{\prime} \mid a_0, \bs{X}^\prime), t)}{\partial \pi} 
\frac{ I(A=a_0)}{\pi^\prime(a_0 \mid  \bs{X}^\prime)}
\pi(s^\prime \mid a_0, \bs{X}^\prime)
S({\pi}(s^{\prime \prime} \mid a_0,  \bs{X}^\prime), t)\, \\
& \cdot r(a_0, s^\prime,  \bs{X}^\prime)\, ds^\prime ds^{\prime \prime}.
\end{align*}

\begin{align*}
&\sum_{s^\prime\in\mathcal{S}} \sum_{s^{\prime \prime}\in\mathcal{S}} \sum_{x^\prime\in\mathcal{X}'} 
K_{h_0}(s^\prime - s_0)K_{h_1}(s^{\prime \prime} - s_1) [(3)] \\
&=\sum_{s^\prime\in\mathcal{S}} \sum_{s^{\prime \prime}\in\mathcal{S}} \sum_{x^\prime\in\mathcal{X}'} 
K_{h_0}(s^\prime - s_0)K_{h_1}(s^{\prime \prime} - s_1)
\frac{\mathbb{I}(\bs{X}^\prime=x^\prime, S=s^\prime, A=a_0)}
{\pi(s^\prime \mid a_0, x^\prime){\pi^\prime}(a_0 \mid  x^\prime)}
\big(Y-r(a_0, s^\prime, x^\prime)\big)\,
S({\pi}(s^{\prime} \mid a_0,  x^\prime), t)\, \\
& \cdot S({\pi}(s^{\prime \prime} \mid a_1,  x^\prime), t)\\
& = \int_{s^{\prime\prime}\in\mathcal{S}} 
K_{h_0}(S - s_0)K_{h_1}(s^{\prime \prime} - s_1)
\frac{\mathbb{I}(A=a_0)}
{\pi(S \mid a_0, \bs{X}^\prime){\pi^\prime}(a_0 \mid  \bs{X}^\prime)}
\big(Y-r(a_0, S, \bs{X}^\prime)\big)\,
S({\pi}(S \mid a_0,  \bs{X}^\prime), t)\, \\
& \cdot S({\pi}(s^{\prime \prime} \mid a_1,  \bs{X}^\prime), t)\, ds^{\prime \prime}.
\end{align*}

\begin{align*}
&\sum_{s^\prime\in\mathcal{S}} \sum_{s^{\prime \prime}\in\mathcal{S}} \sum_{x^\prime\in\mathcal{X}'} 
K_{h_0}(s^\prime - s_0)K_{h_1}(s^{\prime \prime} - s_1) [(4)] \\
& = \sum_{s^\prime\in\mathcal{S}} \sum_{s^{\prime \prime}\in\mathcal{S}} \sum_{x^\prime\in\mathcal{X}'} 
K_{h_0}(s^\prime - s_0)K_{h_1}(s^{\prime \prime} - s_1)
\left[\mathbb{I}(\bs{X}^\prime=x^\prime)-p(x^\prime)\right]\cdot 
S({\pi}(s^{\prime} \mid a_0,  x^\prime), t)\,
S({\pi}(s^{\prime \prime} \mid a_1,  x^\prime), t) \cdot r(a_0, s^\prime,  x^\prime)\\
& = \iint_{\substack{s^\prime\in\mathcal{S},\, s^{\prime\prime}\in\mathcal{S}}}
K_{h_0}(s^\prime - s_0)K_{h_1}(s^{\prime \prime} - s_1) 
S({\pi}(s^{\prime} \mid a_0,  \bs{X}^\prime), t)\,
S({\pi}(s^{\prime \prime} \mid a_1,  \bs{X}^\prime), t) \cdot 
r(a_0, s^\prime,  \bs{X}^\prime)\, ds^\prime ds^{\prime \prime} \\
&\quad - \iiint_{\substack{s^\prime\in\mathcal{S},\, s^{\prime\prime}\in\mathcal{S},\\ x^\prime\in\mathcal{X}'}}
K_{h_0}(s^\prime - s_0)K_{h_1}(s^{\prime \prime} - s_1)
S({\pi}(s^{\prime} \mid a_0,  x^\prime), t)\,
S({\pi}(s^{\prime \prime} \mid a_1,  x^\prime), t)\,
r(a_0, s^\prime,  x^\prime)\, dP(x^\prime)\, ds^\prime ds^{\prime \prime}\\
& = \iint_{\substack{s^\prime\in\mathcal{S},\, s^{\prime\prime}\in\mathcal{S}}}
K_{h_0}(s^\prime - s_0)K_{h_1}(s^{\prime \prime} - s_1)\cdot 
S({\pi}(s^{\prime} \mid a_0,  \bs{X}^\prime), t)\,
S({\pi}(s^{\prime \prime} \mid a_1,  \bs{X}^\prime), t) \cdot 
r(a_0, s^\prime,  \bs{X}^\prime)\, ds^\prime ds^{\prime \prime} \\
&\quad - \tau^{\text{den}}_{\omega_{s_1, s_0}^{\text{sd-trim}}}(a_0, s_0).
\end{align*}
Therefore,

\begin{align*}
    &\mathbb{IF}(\tau^{\text{den}}_{\omega_{s_1, s_0}^{\text{sd-trim}}}(a_0, s_0)) \\
    &=\int_{\substack{s^\prime\in\mathcal{S}}} K_{h_0}(s' - s_0) K_{h_1}(S - s_1)
\frac{I(A = a_1)}{\pi'(a_1 \mid \bs{X}^\prime)}
\frac{\partial S(\pi(S \mid a_1, \bs{X}^\prime), t)}{\partial \pi(S \mid a_1, \bs{X}^\prime)}
S(\pi(s' \mid a_0, \bs{X}^\prime), t) \, r(a_0, s', \bs{X}^\prime) \, ds' \\[8pt]
&\quad +
\int_{\substack{s^{\prime\prime}\in\mathcal{S}}} K_{h_0}(S - s_0) K_{h_1}(s'' - s_1)
\frac{I(A = a_0)}{\pi'(a_0 \mid \bs{X}^\prime)}
\frac{\partial S(\pi(S \mid a_0, \bs{X}^\prime), t)}{\partial \pi(S \mid a_0, \bs{X}^\prime)}
S(\pi(s'' \mid a_1, \bs{X}^\prime), t) \, r(a_0, S, \bs{X}^\prime) \, ds'' \\[8pt]
&\quad +
\int_{\substack{s^{\prime\prime}\in\mathcal{S}}} K_{h_0}(S - s_0) K_{h_1}(s'' - s_1)
\frac{I(A = a_0)}{\pi'(a_0 \mid \bs{X}^\prime) \, \pi(S \mid a_0, \bs{X}^\prime)}
S(\pi(S \mid a_0, \bs{X}^\prime), t) S(\pi(s'' \mid a_1, \bs{X}^\prime), t)\\
& \hspace{3cm} \cdot \big(Y - r(a_0, S, \bs{X}^\prime)\big) \, ds'' \\[8pt]
&\quad -
\iint_{\substack{s^\prime\in\mathcal{S},\, s^{\prime\prime}\in\mathcal{S}}} K_{h_0}(s' - s_0) K_{h_1}(s'' - s_1)
\Bigg[
\frac{I(A = a_1)}{\pi'(a_1 \mid \bs{X}^\prime)}
\frac{\partial S(\pi(s'' \mid a_1, \bs{X}^\prime), t)}{\partial \pi}
\pi(s'' \mid a_1, \bs{X}^\prime)
S(\pi(s' \mid a_0, \bs{X}^\prime), t) \\[4pt]
&\hspace{2.5cm} +
r(a_0, s', \bs{X}^\prime) \frac{I(A = a_0)}{\pi'(a_0 \mid \bs{X}^\prime)}
\frac{\partial S(\pi(s' \mid a_0, \bs{X}^\prime), t)}{\partial \pi}
\pi(s' \mid a_0, \bs{X}^\prime)
S(\pi(s'' \mid a_1, \bs{X}^\prime), t)
r(a_0, s', \bs{X}^\prime)
\Bigg] ds' ds'' \\[8pt]
&\quad +
\iint_{\substack{s^\prime\in\mathcal{S},\, s^{\prime\prime}\in\mathcal{S}}} K_{h_0}(s' - s_0) K_{h_1}(s'' - s_1)
S(\pi(s' \mid a_0, \bs{X}^\prime), t)
S(\pi(s'' \mid a_1, \bs{X}^\prime), t)
r(a_0, s', \bs{X}^\prime) \, ds' ds'' \\[8pt]
&\quad - \, \tau^{\text{den}}_{\omega_{s_1, s_0}^{\text{sd-trim}}}(a_0, s_0).
\end{align*}

Thus, we can derive the efficient influence function for $\tau^{\text{num}}_{\omega_{s_1, s_0}^{\text{sd-trim}}}(a_0, s_0)$ similarly
\begin{align*}
&\mathbb{IF}(\tau^{\text{num}}_{\omega_{s_1, s_0}^{\text{sd-trim}}}(a_1, s_1))\\
&=\int_{\substack{s^\prime\in\mathcal{S}}} K_{h_0}(s' - s_0) K_{h_1}(S - s_1)
\frac{I(A = a_1)}{\pi'(a_1 \mid \bs{X}^\prime)} \frac{\partial S(\pi(S \mid a_1, \bs{X}^\prime), t)}{\partial \pi(S \mid a_1, \bs{X}^\prime)} S(\pi(s' \mid a_0, \bs{X}^\prime), t) \, r(a_1, S, \bs{X}^\prime) \, ds' \\[8pt]
&\quad + \int_{\substack{s^{\prime\prime}\in\mathcal{S}}} K_{h_0}(S - s_0) K_{h_1}(s'' - s_1)
\frac{I(A = a_0)}{\pi'(a_0 \mid \bs{X}^\prime)}
\frac{\partial S(\pi(S \mid a_0, \bs{X}^\prime), t)}{\partial \pi(S \mid a_0, \bs{X}^\prime)}
S(\pi(s'' \mid a_1, \bs{X}^\prime), t) \, r(a_1, s'', \bs{X}^\prime) \, ds'' \\[8pt]
&\quad +
\int_{\substack{s^{\prime\prime}\in\mathcal{S}}} K_{h_0}(S - s_0) K_{h_1}(s'' - s_1)
\frac{I(A = a_1)}{\pi'(a_1 \mid \bs{X}^\prime) \, \pi(S \mid a_1, \bs{X}^\prime)}
S(\pi(S \mid a_0, \bs{X}^\prime), t) S(\pi(S \mid a_1, \bs{X}^\prime), t)\\
& \hspace{3cm}  \cdot \big(Y - r(a_1, S, \bs{X}^\prime)\big) \, ds'' \\[8pt]
&\quad -
\iint_{\substack{s^\prime\in\mathcal{S},\, s^{\prime\prime}\in\mathcal{S}}} K_{h_0}(s' - s_0) K_{h_1}(s'' - s_1)
\Bigg[
\frac{I(A = a_1)}{\pi'(a_1 \mid \bs{X}^\prime)}
\frac{\partial S(\pi(s'' \mid a_1, \bs{X}^\prime), t)}{\partial \pi}
\pi(s'' \mid a_1, \bs{X}^\prime)
S(\pi(s' \mid a_0, \bs{X}^\prime), t) \\[4pt]
&\hspace{3cm} +
\cdot r(a_1, s'', \bs{X}^\prime) \frac{I(A = a_0)}{\pi'(a_0 \mid \bs{X}^\prime)}
\frac{\partial S(\pi(s' \mid a_0, \bs{X}^\prime), t)}{\partial \pi}
\pi(s' \mid a_0, \bs{X}^\prime)
S(\pi(s'' \mid a_1, \bs{X}^\prime), t)
r(a_1, s'', \bs{X}^\prime)
\Bigg] ds' ds'' \\[8pt]
&\quad +
\iint_{\substack{s^\prime\in\mathcal{S},\, s^{\prime\prime}\in\mathcal{S}}} K_{h_0}(s' - s_0) K_{h_1}(s'' - s_1)
S(\pi(s' \mid a_0, \bs{X}^\prime), t)
S(\pi(s'' \mid a_1, \bs{X}^\prime), t)
r(a_1, s'', \bs{X}^\prime) \, ds' ds'' \\[8pt]
&\quad - \,\tau^{\text{num}}_{\omega_{s_1, s_0}^{\text{sd-trim}}}(a_1, s_1).
\end{align*}

\subsection{Analyzing the Remainder Terms of Proposition \ref{prop: EIF for CoR}}
Our estimands are defined as
\begin{equation*} 
\begin{split}
\tau_{\omega_{s}^{\text{s-trim}}}^{\text{num}}(a, s)
&={\iiint_{\substack{s_0\in\mathcal{S},\\ \bs{X}^\prime\in\mathcal{X}'}}
K_h\left(s_0-s\right)\,
S\left(\pi\left(s_0 \mid a, B, \bs{X}\right), t\right)\,
\pi^\prime(a \mid B, \bs{X})\,
r\left(a, s_0, B, \bs{X}\right)\,
d s_0\, d P(B)\, dP(\bs{X})},\\
\tau_{\omega_{s}^{\text{s-trim}}}^{\text{den}}(a, s)
&= {\iiint_{\substack{s_0\in\mathcal{S},\\ \bs{X}^\prime\in\mathcal{X}'}}
K_h\left(s_0-s\right)\,
S\left(\pi\left(s_0 \mid a, B, \bs{X}\right), t\right)\,
\pi^\prime(a \mid B, \bs{X})\,
d s_0\, d P(B)\, dP(\bs{X})}.
\end{split}
\end{equation*}

The centered EIFs are

\begin{equation*} 
\begin{split}
   \varphi^{\text{num}}(a, s) 
   & = \mathbb{IF}\left(\tau_{\omega_{s}^{\text{s-trim}}}^{\text{num}}(a, s)\right) \\
&= K_h(S-s)\,
\frac{\partial S\left(\pi\left(S \mid A, \bs{X}^{\prime}\right), t\right)}
{\partial \pi\left(S \mid A,\bs{X}^{\prime}\right)}\,
r\left(A, S, \bs{X}^{\prime}\right) \\
&\quad +K_h(S-s)\,
\frac{S\left(\pi\left(S \mid A, \bs{X}^{\prime}\right), t\right)}
{\pi\left(S \mid A, \bs{X}^{\prime}\right)}
\left\{Y-r\left(A, S, \bs{X}^{\prime}\right)\right\} \\
&\quad -\int_{s_0\in\mathcal{S}} K_h\left(s_0-s\right)\,
\frac{\partial S\left(\pi\left(s_0 \mid A, \bs{X}^{\prime}\right), t\right)}
{\partial \pi\left(s_0 \mid A, \bs{X}^{\prime}\right)}\,
\pi\left(s_0 \mid A, \bs{X}^{\prime}\right)\,
r\left(A, s_0, \bs{X}^{\prime}\right)\, d s_0 \\
&\quad +\int_{s_0\in\mathcal{S}} K_h\left(s_0-s\right)\,
S\left(\pi\left(s_0 \mid A, \bs{X}^{\prime}\right), t\right)\,
r\left(A, s_0, \bs{X}^{\prime}\right)\, d s_0 
-\tau_{\omega_{s}^{\text{s-trim}}}^{\text{num}}(a, s), \text{ and}
\end{split}
\end{equation*}

\begin{equation*} 
\begin{split}
  \varphi^{\text{den}}(a, s) &=
  \mathbb{IF}\left(\tau_{\omega_{s}^{\text{s-trim}}}^{\text{den}}(a, s)\right) \\
  &=K_h(S-s)\,
  \frac{\partial S\left(\pi\left(S \mid A, \bs{X}^\prime\right), t\right)}
  {\partial \pi\left(S \mid A, \bs{X}^\prime\right)} \\
&\quad -\int_{s_0\in\mathcal{S}} K_h\left(s_0-s\right)\,
\frac{\partial S\left(\pi\left(s_0 \mid A, \bs{X}^\prime\right), t\right)}
{\partial \pi\left(s_0 \mid A, \bs{X}^\prime\right)}\,
\pi\left(s_0 \mid A, \bs{X}^\prime\right)\, d s_0 \\
&\quad + \int_{s_0\in\mathcal{S}} K_h\left(s_0-s\right)\,
S\left(\pi\left(s_0 \mid A, \bs{X}^\prime\right), t\right)\, d s_0 
-\tau_{\omega_{s}^{\text{s-trim}}}^{\text{den}}(a, s).
\end{split}
\end{equation*}
For completeness, we'll first confirm that $\mathbb{E}\left[ \varphi^{\text{den}}(a, s)\right]=0$ and $\mathbb{E}\left[ \varphi^{\text{num}}(a, s)\right]=0$ in order for the von Mises expansion to be valid.

First let's start with $\varphi^{\text{den}}(a, s)$. Note that

\begin{equation*}
\begin{split}
&\mathbb{E}\left[K_h(S-s)\,
\frac{\partial S\left(\pi\left(S \mid A, \bs{X}^\prime\right), t\right)}
{\partial \pi\left(S \mid A, \bs{X}^\prime\right)}\right]
=\iiint_{\substack{s_0\in\mathcal{S},\, a\in\mathcal{A},\\ x^\prime\in\mathcal{X}'}}
K_h(s_0-s)\,
\frac{\partial S\left(\pi\left(s_0 \mid a, x^\prime\right), t\right)}
{\partial \pi\left(s_0 \mid a, x^\prime\right)}
\cdot \pi(s_0 \mid a, x^\prime)\cdot \pi^\prime(a \mid x^\prime)\,
ds_0\, da\, dP(x^\prime),\\
&\mathbb{E}\left[\int_{s_0\in\mathcal{S}} K_h\left(s_0-s\right)
\frac{\partial S\left(\pi\left(s_0 \mid A, \bs{X}^\prime\right), t\right)}
{\partial \pi\left(s_0 \mid A, \bs{X}^\prime\right)}
\pi\left(s_0 \mid A, \bs{X}^\prime\right)\, d s_0 \right] \\
&\qquad= \iiint_{\substack{s_0\in\mathcal{S},\, a\in\mathcal{A},\\ x^\prime\in\mathcal{X}'}}
K_h\left(s_0-s\right)
\frac{\partial S\left(\pi\left(s_0 \mid a, x^\prime\right), t\right)}
{\partial \pi\left(s_0 \mid a, x^\prime\right)}
\pi\left(s_0 \mid a, x^\prime\right)\cdot \pi^\prime(a \mid x^\prime)\,
d s_0\, da\, dP(x^\prime),\\
& \mathbb{E}\left[\int_{s_0\in\mathcal{S}} K_h\left(s_0-s\right)
S\left(\pi\left(s_0 \mid A, \bs{X}^\prime\right), t\right)\, d s_0\right] \\
&\qquad= \iiint_{\substack{s_0\in\mathcal{S},\, a\in\mathcal{A},\\ x^\prime\in\mathcal{X}'}}
K_h\left(s_0-s\right)
S\left(\pi\left(s_0 \mid a, x^\prime\right), t\right)\,
\pi^\prime(a \mid x^\prime)\,
d s_0\, da\, dP(x^\prime)
= \tau_{\omega_{s}^{\text{s-trim}}}^{\text{den}}(a, s).
\end{split}
\end{equation*}
Thus, we see that $\mathbb{E}\left[ \varphi^{\text{den}}(a, s)\right]=0$.

For $\varphi^{\text{den}}(a, s)$, note that the expectation of the first term is

\begin{equation*}
\begin{split}
&\mathbb{E}\left[K_h(S-s)\,
\frac{\partial S\left(\pi\left(S \mid A, \bs{X}^{\prime}\right), t\right)}
{\partial \pi\left(S \mid A,\bs{X}^{\prime}\right)}\,
r\left(A, S, \bs{X}^{\prime}\right)\right]\\
&\qquad= \iiint_{\substack{s_0\in\mathcal{S},\, a\in\mathcal{A},\\ x^{\prime}\in\mathcal{X}'}}
K_h(s_0-s)\,
\frac{\partial S\left(\pi\left(s_0 \mid a, x^{\prime}\right), t\right)}
{\partial \pi\left(s_0 \mid a,x^{\prime}\right)}\,
r\left(a, s_0, x^{\prime}\right)\cdot
\pi(s_0\mid a, x^\prime)\cdot \pi^\prime(a \mid x^\prime)\,
ds_0\, da\, dP(x^\prime), \\[3pt]
&\text{and the expectation of the third term is}\\
&\mathbb{E}\left[\int_{s_0\in\mathcal{S}} K_h\left(s_0-s\right)
\frac{\partial S\left(\pi\left(s_0 \mid A, \bs{X}^{\prime}\right), t\right)}
{\partial \pi\left(s_0 \mid A, \bs{X}^{\prime}\right)}
\pi\left(s_0 \mid A, \bs{X}^{\prime}\right)
r\left(A, s_0, \bs{X}^{\prime}\right)\, d s_0 \right]\\
&\qquad=\iiint_{\substack{s_0\in\mathcal{S},\, a\in\mathcal{A},\\ x^{\prime}\in\mathcal{X}'}}
K_h\left(s_0-s\right)
\frac{\partial S\left(\pi\left(s_0 \mid a, x^{\prime}\right), t\right)}
{\partial \pi\left(s_0 \mid a, x^{\prime}\right)}
\pi\left(s_0 \mid a, x^{\prime}\right)
r\left(a, s_0, x^{\prime}\right)\cdot \pi^\prime(a\mid x^\prime)\,
d s_0\, da\, dP(x^\prime).
\end{split}
\end{equation*}
which is exactly the same with the expection of the first term.

Meanwhile, the expectation of the fourth term of $\varphi^{\text{den}}(a, s)$ is

\begin{equation*}
\begin{split}
&\mathbb{E}\left[\int_{s_0\in\mathcal{S}} K_h\left(s_0-s\right)
S\left(\pi\left(s_0 \mid A, \bs{X}^{\prime}\right), t\right)
r\left(A, s_0, \bs{X}^{\prime}\right)\, d s_0 \right]\\
&\qquad = \iiint_{\substack{s_0\in\mathcal{S},\, a\in\mathcal{A},\\ x^{\prime}\in\mathcal{X}'}}
K_h\left(s_0-s\right)
S\left(\pi\left(s_0 \mid a, x^{\prime}\right), t\right)
r\left(a, s_0, x^{\prime}\right)\,
\pi^\prime(a \mid x^\prime)\,
d s_0\, da\, dP(x^\prime)
=  \tau_{\omega_{s}^{\text{s-trim}}}^{\text{num}}(a, s).
\end{split}
\end{equation*}

By the definition $r\left(A, S, \bs{X}^{\prime}\right) = \mathbb{E}[Y(A, S) \mid \bs{X}^\prime]$, we have

\begin{equation*}
\begin{split}
&\mathbb{E}\left[K_h(S-s)\cdot 
\frac{S\left(\pi\left(S \mid A, \bs{X}^{\prime}\right), t\right)}
{\pi\left(S \mid A, \bs{X}^{\prime}\right)}
\left(Y-r\left(A, S, \bs{X}^{\prime}\right)\right)\right]\\
&\quad = \mathbb{E}\Bigg[\mathbb{E}\Bigg[\Bigg.
K_h(S-s)\cdot 
\frac{S\left(\pi\left(S \mid A, \bs{X}^{\prime}\right), t\right)}
{\pi\left(S \mid A, \bs{X}^{\prime}\right)}
\left(Y-r\left(A, S, \bs{X}^{\prime}\right)\right)
\ \Bigg|\ S\in\mathcal{S},\, A\in\mathcal{A},\, \bs{X}^\prime\in\mathcal{X}'\Bigg]\Bigg] \\
&\quad=\mathbb{E}\left[K_h(S-s)\cdot 
\frac{S\left(\pi\left(S \mid A, \bs{X}^{\prime}\right), t\right)}
{\pi\left(S \mid A, \bs{X}^{\prime}\right)}
\left(\mathbb{E}\!\left[Y\mid S\in\mathcal{S},\,A\in\mathcal{A},\,\bs{X}^\prime\in\mathcal{X}'\right]
-r\left(A, S, \bs{X}^{\prime}\right)\right)\right]\\
&\quad=\mathbb{E}\left[K_h(S-s)\cdot 
\frac{S\left(\pi\left(S \mid A, \bs{X}^{\prime}\right), t\right)}
{\pi\left(S \mid A, \bs{X}^{\prime}\right)}
\left(r\left(A, S, \bs{X}^{\prime}\right)-r\left(A, S, \bs{X}^{\prime}\right)\right)\right]\\
&\quad=0.
\end{split}
\end{equation*}

Thus, we see that $\mathbb{E}\left[ \varphi^{\text{num}}(a, s)\right]=0$.

\section{Cross-fitting EIF-based estimators}
\label{app:crossfitting}

This section describes the cross-fitting procedure used to construct the one-step estimators
for the smoothed and trimmed parameters underlying $\mathrm{STWCR}(a,s)$ and
$\mathrm{STWCRVE}(s_1,s_0)$. Throughout, we follow the cross-fitting framework of
\citet{chernozhukov2018double}.

Let $O_i=(Y_i,A_i,S_i,B_i,\mathbf{X}_i)$, $i=1,\ldots,n$, be an i.i.d.\ sample from $P_0$.
Fix an integer $K\ge 2$ and let $\{I_k\}_{k=1}^K$ be a random partition of $\{1,\ldots,n\}$
into $K$ folds of (approximately) equal size. Define the training indices
$I_k^c=\{1,\ldots,n\}\setminus I_k$.
For any measurable function $f$, let
\[
\mathbb{P}_{n,k} f
=
\frac{1}{|I_k|}
\sum_{i\in I_k} f(O_i),
\qquad
\mathbb{P}_{n,k}^c f
=
\frac{1}{|I_k^c|}
\sum_{i\in I_k^c} f(O_i).
\]
We estimate nuisance functions on the training folds $I_k^c$ and evaluate influence-function
expressions on the held-out folds $I_k$.

\subsection{Cross-fitted estimation of $\widehat{\tau}(a,s)$}
\label{app:crossfitting-tau}

Proposition~\ref{prop: EIF for CoR} provides the uncentered efficient influence functions (EIFs)
$\varphi^{\mathrm{num}}(a,s)$ and $\varphi^{\mathrm{den}}(a,s)$ for the numerator and denominator
functionals
$\tau_{\omega_s^{\mathrm{s\text{-}trim}}}^{\mathrm{num}}(a,s)$ and
$\tau_{\omega_s^{\mathrm{s\text{-}trim}}}^{\mathrm{den}}(a,s)$, respectively, under fixed
smoothing parameters $(t,h,\epsilon)$.

\paragraph{Step I (fold-specific nuisance estimation).}
For each fold $k\in\{1,\ldots,K\}$, estimate the nuisance components
\[
\pi'(a\mid B,\mathbf{X}) = P(A=a\mid B,\mathbf{X}),
\pi(s\mid a,B,\mathbf{X}) = P(S=s\mid A=a,B,\mathbf{X}),
r(a,s,B,\mathbf{X}) = \mathbb{E}(Y\mid A=a,S=s,B,\mathbf{X}),
\]
using only the training sample $\{O_i:i\in I_k^c\}$, yielding fold-specific estimators
\[
\widehat{\pi}'_{n,k}(a\mid B,\mathbf{X}),
\widehat{\pi}_{n,k}(s\mid a,B,\mathbf{X}),
\widehat{r}_{n,k}(a,s,B,\mathbf{X}).
\]
In our implementation, each nuisance component is estimated via Super Learner, using a
prespecified library of candidate learners.

\paragraph{Step II (fold-specific EIF evaluation).}
Let $\widehat{\varphi}^{\mathrm{num}}_{k}(a,s)$ and $\widehat{\varphi}^{\mathrm{den}}_{k}(a,s)$
denote the EIFs in Proposition~\ref{prop: EIF for CoR} with nuisance components replaced by their
fold-specific estimates:
\[
\widehat{\varphi}^{\mathrm{num}}_{k}(a,s)
=
\varphi^{\mathrm{num}}\!\left(a,s;
\widehat{\pi}'_{n,k},\widehat{\pi}_{n,k},\widehat{r}_{n,k}\right),
\qquad
\widehat{\varphi}^{\mathrm{den}}_{k}(a,s)
=
\varphi^{\mathrm{den}}\!\left(a,s;
\widehat{\pi}'_{n,k},\widehat{\pi}_{n,k},\widehat{r}_{n,k}\right).
\]

\paragraph{Step III (cross-fitted one-step estimators).}
Define the cross-fitted one-step estimators of the numerator and denominator as
\begin{equation}
\label{eq:tau-num-den-cf}
\widehat{\tau}_{\omega_s^{\mathrm{s\text{-}trim}}}^{\mathrm{num}}(a,s)
=
\frac{1}{K}\sum_{k=1}^K \mathbb{P}_{n,k}\!\left\{\widehat{\varphi}^{\mathrm{num}}_{k}(a,s)\right\},
\qquad
\widehat{\tau}_{\omega_s^{\mathrm{s\text{-}trim}}}^{\mathrm{den}}(a,s)
=
\frac{1}{K}\sum_{k=1}^K \mathbb{P}_{n,k}\!\left\{\widehat{\varphi}^{\mathrm{den}}_{k}(a,s)\right\}.
\end{equation}
The estimator of $\mathrm{STWCR}(a,s)$ is the ratio
\begin{equation}
\label{eq:tau-cf}
\widehat{\tau}(a,s)
=
\frac{
\widehat{\tau}_{\omega_s^{\mathrm{s\text{-}trim}}}^{\mathrm{num}}(a,s)
}{
\widehat{\tau}_{\omega_s^{\mathrm{s\text{-}trim}}}^{\mathrm{den}}(a,s)
}.
\end{equation}
By construction, the held-out fold used in $\mathbb{P}_{n,k}\{\cdot\}$ is independent of the
training fold used to build $\widehat{\pi}'_{n,k}$, $\widehat{\pi}_{n,k}$, and
$\widehat{r}_{n,k}$, which removes first-order overfitting bias and yields the asymptotic
properties in Proposition~\ref{prop: asymp-CoR} under the stated rate conditions.

\subsection{Cross-fitted estimation of $\widehat{\delta}(a_1, a_0, s_1,s_0)$}
\label{app:crossfitting-delta}

Theorem~\ref{thm: EIF CVE} provides the uncentered EIFs
$\theta^{\mathrm{num}}(s_1,s_0)$ and $\theta^{\mathrm{den}}(a_0,s_0)$ for the numerator and
denominator functionals
$\tau_{\omega_{s_1,s_0}^{\mathrm{sd\text{-}trim}}}^{\mathrm{num}}(a_1,s_1)$ and
$\tau_{\omega_{s_1,s_0}^{\mathrm{sd\text{-}trim}}}^{\mathrm{den}}(a_0,s_0)$, respectively, under
fixed smoothing parameters $(t,h_0,h_1,\epsilon)$.

\paragraph{Step I (fold-specific nuisance estimation).}
Using the same $K$-fold partition $\{I_k\}_{k=1}^K$, for each fold $k$ estimate the nuisance
components $\pi'(a\mid B,\mathbf{X})$, $\pi(s\mid a,B,\mathbf{X})$, and $r(a,s,B,\mathbf{X})$
on the training sample $\{O_i:i\in I_k^c\}$, yielding the fold-specific estimators
$\widehat{\pi}'_{n,k}$, $\widehat{\pi}_{n,k}$, and $\widehat{r}_{n,k}$.
The same Super Learner library is used as in Appendix~\ref{app:crossfitting-tau}.

\paragraph{Step II (fold-specific EIF evaluation).}
Define the plug-in EIF estimators
\[
\widehat{\theta}^{\mathrm{num}}_{k}(a_1,s_1)
=
\theta^{\mathrm{num}}\!\left(a_1,s_1;
\widehat{\pi}'_{n,k},\widehat{\pi}_{n,k},\widehat{r}_{n,k}\right),
\qquad
\widehat{\theta}^{\mathrm{den}}_{k}(a_0,s_0)
=
\theta^{\mathrm{den}}\!\left(a_0,s_0;
\widehat{\pi}'_{n,k},\widehat{\pi}_{n,k},\widehat{r}_{n,k}\right),
\]
obtained by replacing the nuisance components in Theorem~\ref{thm: EIF CVE} with their
fold-specific estimates.

\paragraph{Step III (cross-fitted one-step estimators).}
Define the cross-fitted one-step estimators of the numerator and denominator as
\begin{equation}
\label{eq:delta-num-den-cf}
\widehat{\tau}_{\omega_{a_1,s_1}^{\mathrm{sd\text{-}trim}}}^{\mathrm{num}}(a_1,s_1)
=
\frac{1}{K}\sum_{k=1}^K \mathbb{P}_{n,k}\!\left\{\widehat{\theta}^{\mathrm{num}}_{k}(a_1,s_1)\right\},
\qquad
\widehat{\tau}_{\omega_{s_1,s_0}^{\mathrm{sd\text{-}trim}}}^{\mathrm{den}}(a_0,s_0)
=
\frac{1}{K}\sum_{k=1}^K \mathbb{P}_{n,k}\!\left\{\widehat{\theta}^{\mathrm{den}}_{k}(a_0,s_0)\right\}.
\end{equation}
The cross-fitted estimator of $\mathrm{STWCRVE}(a_1, a_0, s_1,s_0)$ is then
\begin{equation}
\label{eq:delta-cf}
\widehat{\delta}(a_1,s_1, a_0, s_0)
=
1-
\frac{
\widehat{\tau}_{\omega_{s_1,s_0}^{\mathrm{sd\text{-}trim}}}^{\mathrm{num}}(a_1,s_1)
}{
\widehat{\tau}_{\omega_{s_1,s_0}^{\mathrm{sd\text{-}trim}}}^{\mathrm{den}}(a_0,s_0)
}.
\end{equation}
Cross-fitting again ensures that EIF evaluation is performed on data independent of nuisance
estimation. Under the same conditions as Proposition~\ref{prop: asymp-CoR}, with $(a,s)$ replaced
by $(s_1,s_0)$ and $(\varphi^{\mathrm{num}},\varphi^{\mathrm{den}})$ replaced by
$(\theta^{\mathrm{num}},\theta^{\mathrm{den}})$, Theorem~\ref{thm: EIF_CVE_property} yields
asymptotic normality and provides a consistent variance estimator based on the empirical variance
of the corresponding estimated influence function.

\section{Regularity conditions}
\label{app: regularity condition}

\subsection{Regularity conditions for Proposition \ref{prop: asymp-CoR}}
\label{app: regularity condition for prop cor}
\begin{enumerate}[(i)]
\item \textbf{Independent-sample (cross-fitting) assumption.}  
The folds used to estimate $\widehat{\pi}^\prime$, $\widehat{\pi}(\cdot)$, and $\widehat{r}$ are independent of 
the folds used to evaluate $\widehat{\varphi}^{\mathrm{num}}(a,s)$ and $\widehat{\varphi}^{\mathrm{den}}(a,s)$.

\item \textbf{Rate conditions.} The nuisance estimators satisfy
\[
\|\widehat{\pi}(s\mid a,B,\mathbf{X}) - \pi(s\mid a,B,\mathbf{X})\|
 = o_p(n^{-1/4}),\qquad
\|\widehat{\pi}^\prime(a\mid B,\mathbf{X}) - \pi^\prime(a\mid B,\mathbf{X})\|
 = o_p(n^{-1/4}),
\]
and
\[
\|\widehat{r}(a,s,B,\mathbf{X}) - r(a,s,B,\mathbf{X})\|
 = o_p(n^{-1/4}),
\]
where the norm denotes $L_2(P)$.  
These conditions ensure that the remainder terms of the von Mises expansion of the EIFs are $o_p(n^{-1/2})$.

\item \textbf{Consistency of EIF estimators.}  
\[
\|\widehat{\varphi}^{\mathrm{num}}(a,s)-\varphi^{\mathrm{num}}(a,s)\| = o_p(1), 
\qquad
\|\widehat{\varphi}^{\mathrm{den}}(a,s)-\varphi^{\mathrm{den}}(a,s)\| = o_p(1).
\]
\end{enumerate}
\subsection{Regularity conditions for Theorem \ref{thm: EIF_CVE_property}}
\label{app: regularity condition for thm cve}

\begin{enumerate}[(i)]
\item \textbf{Independent-sample (cross-fitting) assumption.}  
The folds used to estimate all nuisance functions appearing in 
$\widehat{\theta}^{\mathrm{num}}(a_1,s_1)$ and $\widehat{\theta}^{\mathrm{den}}(a_0,s_0)$
(e.g., $\widehat{\pi}^\prime$, $\widehat{\pi}(\cdot)$, $\widehat{r}$ and any additional nuisance components required by $\theta^{\mathrm{num}}$ and $\theta^{\mathrm{den}}$)
are independent of the folds used to evaluate the cross-fitted EIF estimates 
$\widehat{\theta}^{\mathrm{num}}(a_1,s_1)$ and $\widehat{\theta}^{\mathrm{den}}(a_0,s_0)$.

\item \textbf{Rate conditions.}  
The nuisance estimators satisfy 
\[
\|\widehat{\pi}(s\mid a,B,\mathbf{X}) - \pi(s\mid a,B,\mathbf{X})\|
 = o_p(n^{-1/4}),\qquad
\|\widehat{\pi}^\prime(a\mid B,\mathbf{X}) - \pi^\prime(a\mid B,\mathbf{X})\|
 = o_p(n^{-1/4}),
\]
and
\[
\|\widehat{r}(a,s,B,\mathbf{X}) - r(a,s,B,\mathbf{X})\|
 = o_p(n^{-1/4}),
\]
where the norm denotes $L_2(P)$. These conditions ensure that the von Mises expansion remainder terms for the EIFs are
$o_p(n^{-1/2})$.

\item \textbf{Consistency of EIF estimators.}  
\[
\|\widehat{\theta}^{\mathrm{num}}(a_1,s_1)-\theta^{\mathrm{num}}(a_1,s_1)\| = o_p(1), 
\qquad
\|\widehat{\theta}^{\mathrm{den}}(a_0,s_0)-\theta^{\mathrm{den}}(a_0,s_0)\| = o_p(1).
\]
\end{enumerate}

\section{Additional simulation results}
\label{app: add simu rslts}
\begin{figure}[H]
  \centering
  \begin{minipage}[b]{0.45\textwidth}
    \centering
    \includegraphics[width=\textwidth]{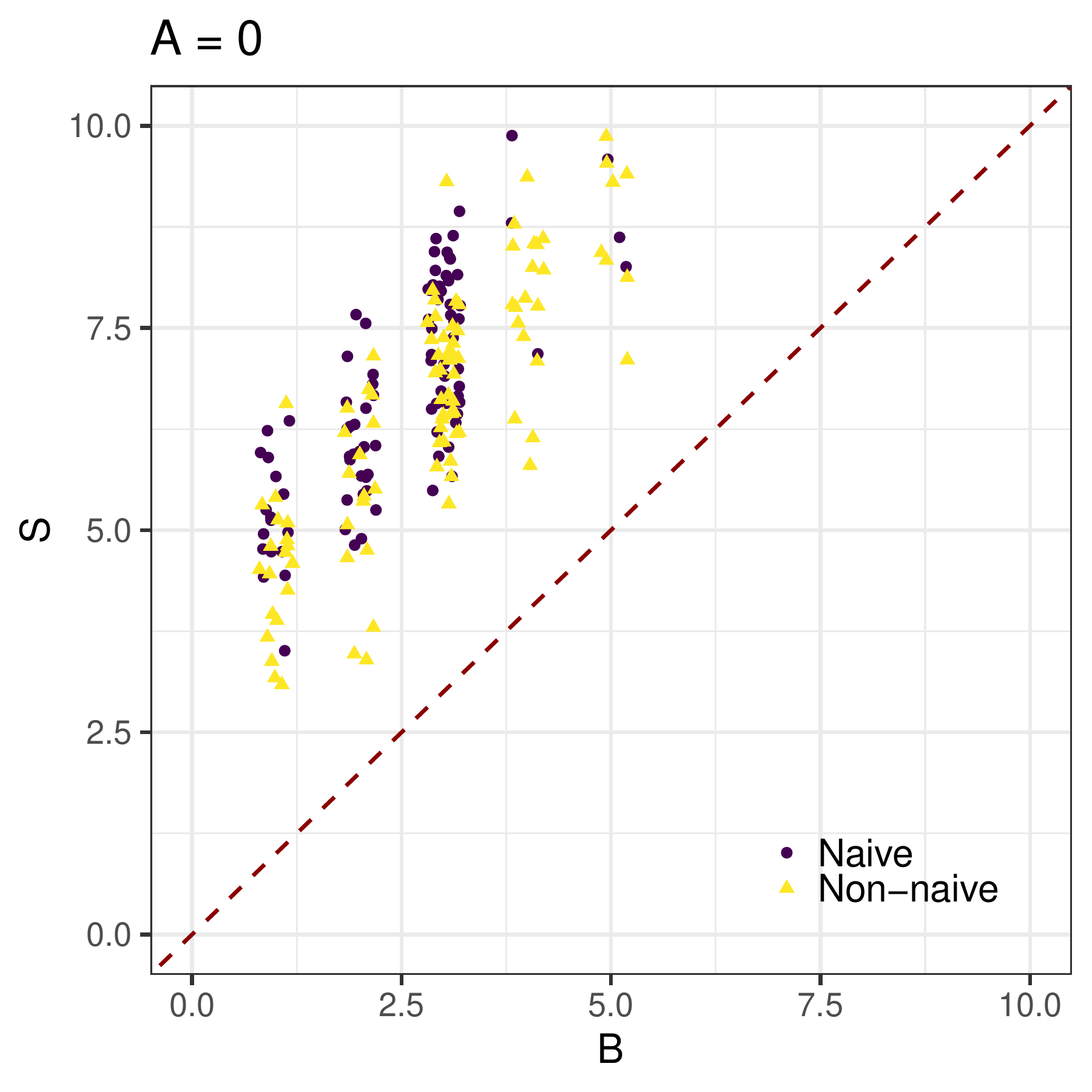}
  \end{minipage}
  \hfill
  \begin{minipage}[b]{0.45\textwidth}
    \centering
    \includegraphics[width=\textwidth]{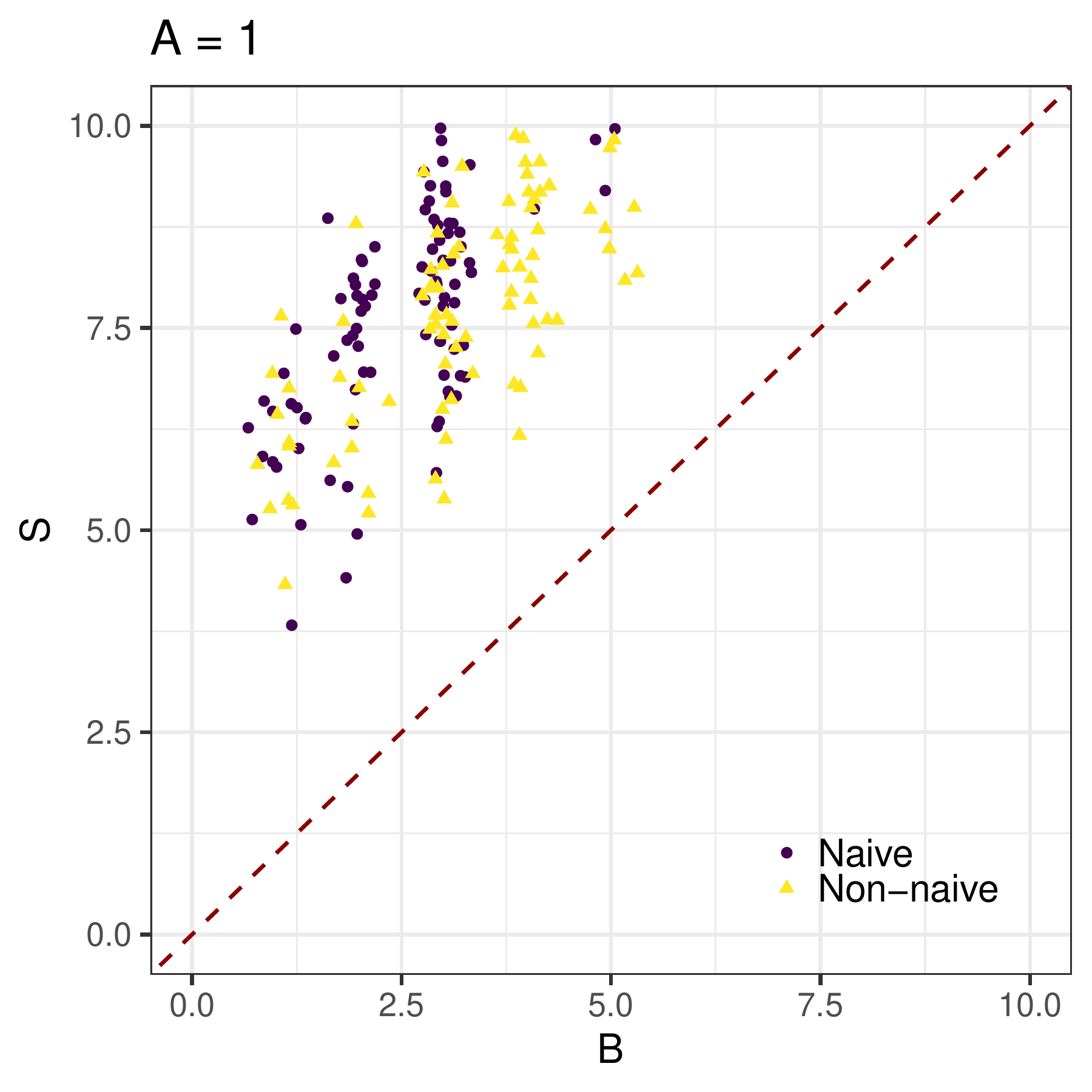} 
      \end{minipage} \\
      \begin{minipage}[b]{0.45\textwidth}
    \centering
    \includegraphics[width=\textwidth]{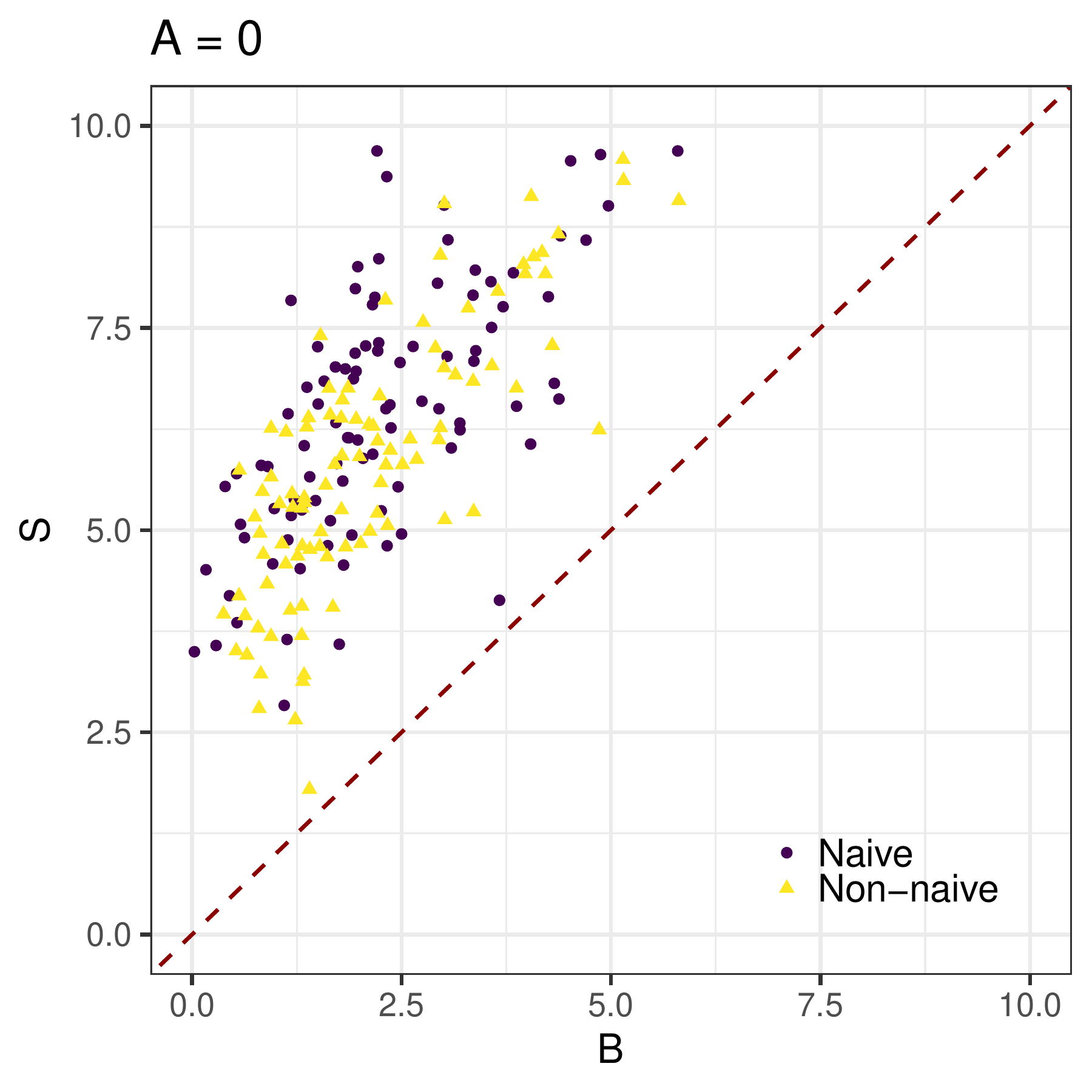}
   
  \end{minipage}
  \hfill
  \begin{minipage}[b]{0.45\textwidth}
    \centering
    \includegraphics[width=\textwidth]{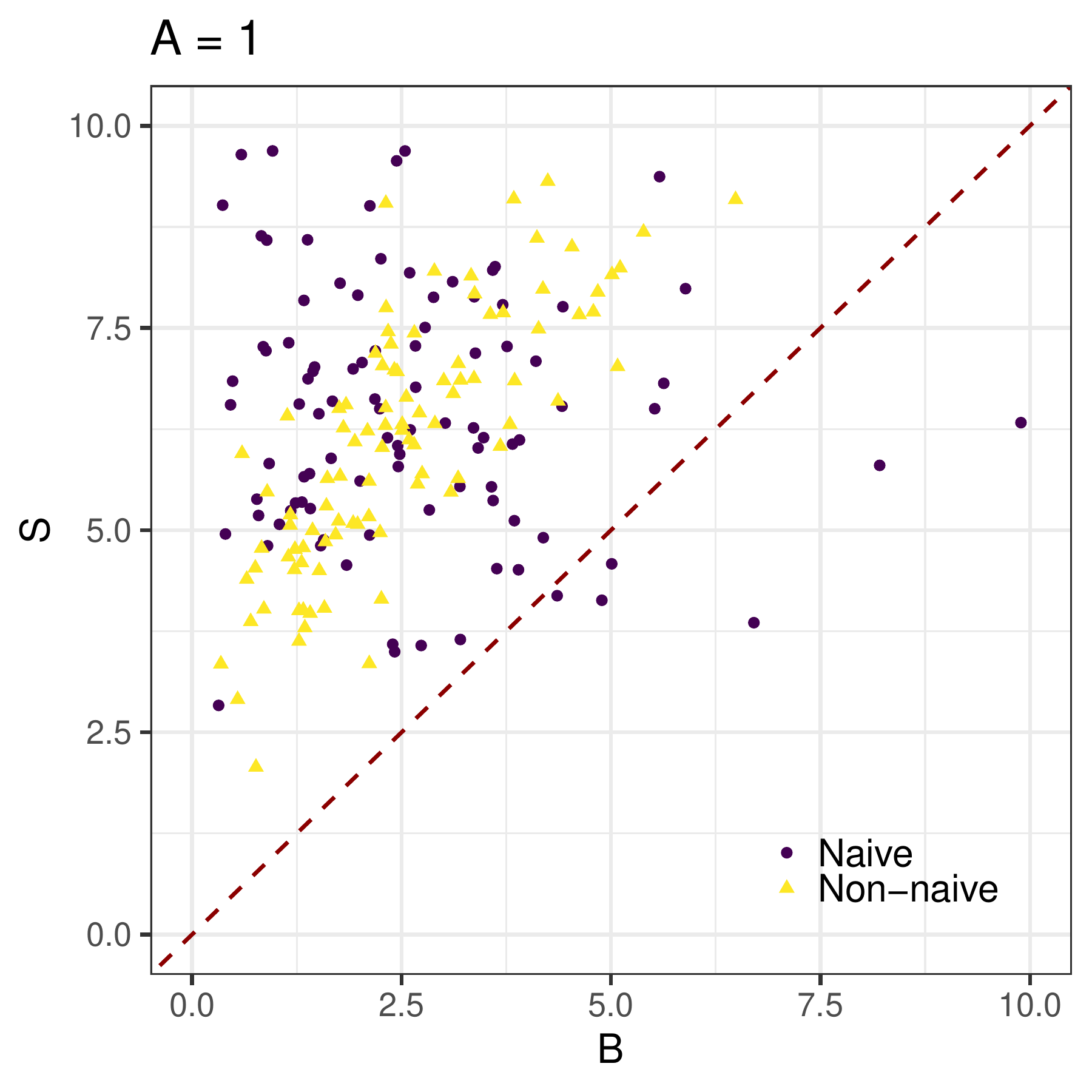}
      \end{minipage}
  \caption{An illustration of the joint distribution of the baseline immune marker level $B$ and the peak level $S$ among na\"ive (yellow) and non-na\"ive (purple) participants under Factor 2 Scenario I (top panels) and Scenario II (bottom panels). A small horizontal jitter was added for exhibition purpose in Scenario I. Left panels correspond to participants receiving the comparator vaccine ($A=0$), and right panels correspond to participants receiving the investigational vaccine ($A=1$).}
  \label{fig: simu S vs B}
\end{figure}

\begin{figure}[H]
    \centering
    \includegraphics[width=\linewidth]{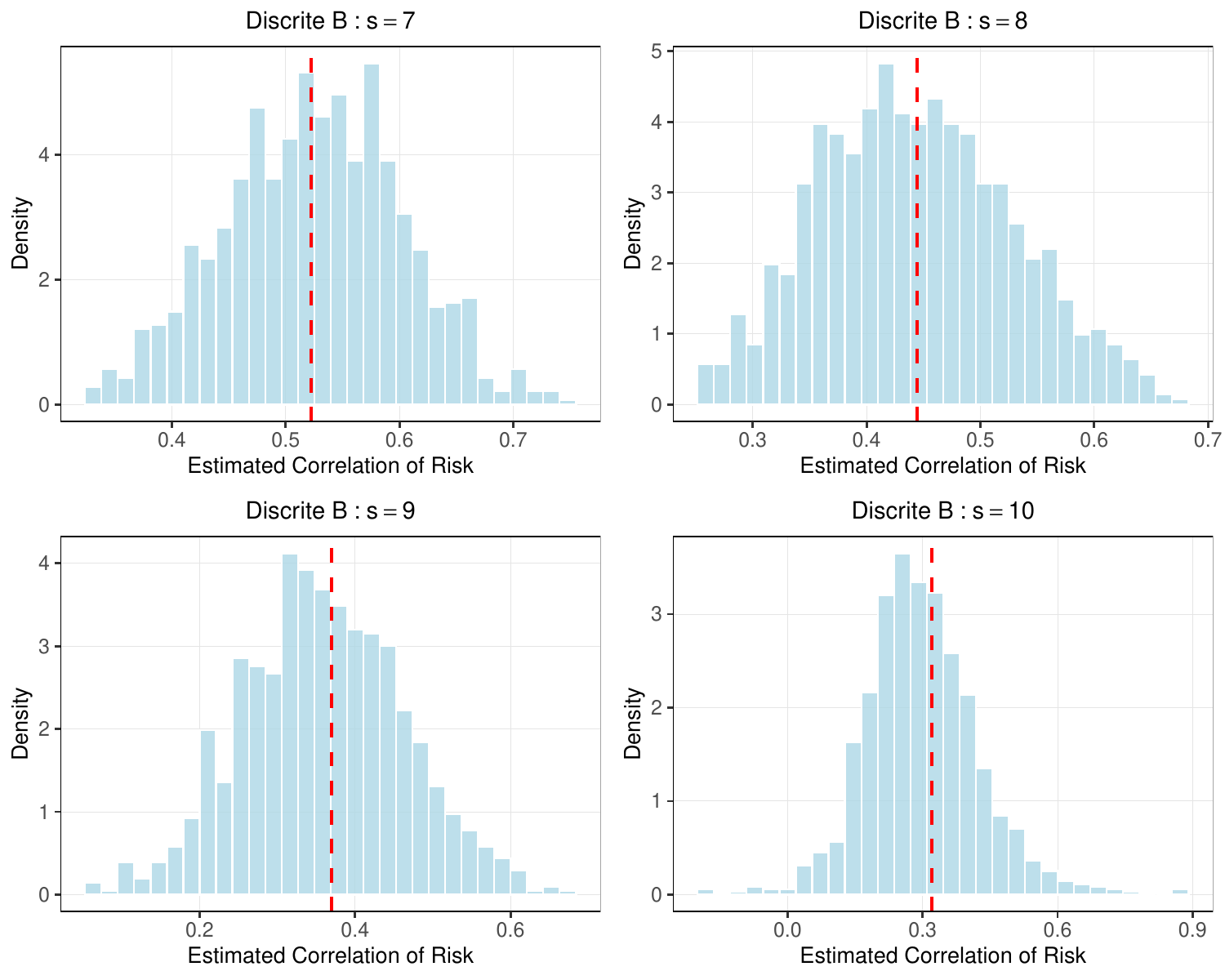}
    \caption{Sampling distributions of EIF-based estimator for STWCR(1, s) when B is continuous (\textsf{Scenario II}). The sample size is 1000 and simulations are repeated for 1000 times. The dashed red lines represent the true STWCR(1, s) for each $s$.}
    \label{fig:EIF distribution B cont}
\end{figure}

\newpage
\section{Additional details on the case study}
\label{app: case study}

\begin{table}[h!]
\centering
\caption{Estimated STWCRVE$(1, 0, s_1, s_0)$ under different $s_0$ and $s_1$.}
\label{tab:cve_s0_s1}
\begin{tabular}{ccccccc}
\hline
$s_0$ & $s_1$ & $\widehat{\delta}(1, 0, s_1, s_0)$ & $\widehat{\tau}_{\omega_{s_1, s_0}^{\text{sd-trim}}}^{\text{num}}(1, s_1)$ & $\widehat{\tau}_{\omega_{s_1, s_0}^{\text{sd-trim}}}^{\text{den}}(0, s_0)$ & SE & 95\% CI \\
\hline
3.50 & 3.75 & $24.2\%$  & $33.1\%$ & $43.7\%$ & $0.203$ & $(-28.0\%,\; 55.2\%)$ \\
3.50 & 4.00 & $28.1\%$  & $27.4\%$ & $38.1\%$ & $0.255$ & $(-44.1\%,\; 64.1\%)$ \\
3.75 & 3.50 & $-33.3\%$ & $62.2\%$ & $46.7\%$ & $0.249$ & $(-92.4\%,\; 7.6\%)$ \\
3.75 & 4.00 & $27.7\%$  & $28.8\%$ & $39.8\%$ & $0.196$ & $(-23.0\%,\; 57.5\%)$ \\
4.00 & 3.50 & $-191.5\%$& $55.5\%$ & $19.0\%$ & $2.085$ & $(-1084.7\%,\; 28.3\%)$ \\
4.00 & 3.75 & $-77.6\%$ & $41.0\%$ & $23.1\%$ & $0.780$ & $(-320.0\%,\; 24.9\%)$ \\
\hline
\end{tabular}
\end{table}